%% file: report7_long.tex
\documentclass[journal,draftclsnofoot,onecolumn,12pt]{IEEEtran}
\usepackage{amsthm,amssymb,graphicx,multirow,amsmath,color,amsfonts}
\usepackage[update,prepend]{epstopdf}
\usepackage[noadjust]{cite}
\usepackage[latin1,utf8]{inputenc}
\usepackage{tikz}
\usepackage{bbm} 
\usepackage{pdfpages}
\usepackage{tabulary}
\usepackage{multirow}
\usepackage{comment}
\usepackage{subfigure}
\usepackage{float}

\DeclareUnicodeCharacter{0177}{\^y}

\include{notation}

\allowdisplaybreaks 
\begin{document}
\title{
	 A Dominant Interferer plus Mean Field-based Approximation for SINR Meta Distribution in Wireless Networks
}
\author{
	Yujie Qin, Mustafa A. Kishk, {\em Member, IEEE}, and Mohamed-Slim Alouini, {\em Fellow, IEEE}
	\thanks{Yujie Qin and Mohamed-Slim Alouini are with Computer, Electrical and Mathematical Sciences and Engineering (CEMSE) Division, King Abdullah University of Science and Technology (KAUST), Thuwal, 23955-6900, Saudi Arabia
		Arabia. Mustafa Kishk is with the Department of Electronic Engineering, Maynooth University, Maynooth, W23 F2H6, Ireland. (e-mail: yujie.qin@kaust.edu.sa; mustafa.kishk@mu.ie; slim.alouini@kaust.edu.sa).} 
	
}
\date{\today}
\maketitle

\begin{abstract}
	This paper proposes a novel approach for computing the meta distribution of the signal-to-interference-plus-noise ratio (SINR) for the downlink transmission in a wireless network with Rayleigh fading.  The novel approach relies on an approximation mix of exact and mean-field analysis of interference (dominant interferer-based approximation) to reduce the complexity of analysis and enhance tractability. In particular, the proposed approximation omits the need to compute the first or the second moment of the SINR that is used in the beta approximation typically adopted in the literature but requires of computing the joint distance distributions. We first derive the proposed approximation based on a Poisson point process (PPP) network with a standard path-loss and Rayleigh fading and then illustrate its accuracy and operability in another four widely used point processes: Poisson bipolar network, Mat\'{e}rn cluster process (MCP), $K$-tier PPP and Poisson line Cox process (PLCP).  Specifically, we obtain the SINR meta distribution for PLCP networks for the first time. Even though the proposed approximation looks simple but it shows good matching in comparison to the popular beta approximation as well as the Monte-Carlo simulations, which opens the door to adopting this approximation in more advanced network architectures.
\end{abstract}

\section{Introduction}

\subsection{Motivation}
In wireless communication, the accurate modeling of the locations of the base stations (BSs) is essential to characterize the system performance and obtain the critical design insights \cite{haenggi2009stochastic}. Traditionally, the locations of BSs are modeled by lattices, which are intractable in analysis \cite{win2009mathematical}. However, to enhance the spatial reuse and meet an exponential growth in mobile traffic, the deployment of BSs becomes irregular and heterogeneous. For instance, macro, pico, and femto BSs can coexist, and UAVs or other high altitude platforms are deployed to help offload the ground BSs, which yield different path loss exponents \cite{singh2013offloading,alzenad2019coverage} and more complicated system models. In this case, stochastic geometry provides the tools which are widely utilized in modeling, characterizing, and obtaining design insights of the wireless networks with randomly placed nodes \cite{haenggi2015meta,elsawy2017meta}. Among many point processes, the Poisson point process (PPP) is a widely used model due to its analytical tractability and  stationarity. This property results in a simple expression for the probability generating functional (PGFL) \cite{andrews2016primer}. While most of the stochastic geometry-based analyses are confined to the spatial average, the performance from the perspective of each user is ignored.  Taking coverage probability, for example, this performance is obtained by averaging over the channel fading and the point process by utilizing the Laplace transform of the interference with the aid of PGFL \cite{elsawy2016modeling}. Such performance metric quantifies the overall  signal-to-interference-plus-noise ratio (SINR) performance, however, limited information about the individual links. For example, a cellular network with a coverage probability of $0.8$ does not mean that all the links have the same success probability, some may have $0.95$ while the others may have $0.6$, and two networks with the same coverage probabilities would differ greatly. In other words,  it is crucial to obtain key information about "the distribution of success probability of the individual link in a given network" \cite{kalamkar2019simple}, which reveals the reliability and quality of service (QoS) of the network and is a fundamental design objective for cellular operators. This new and fundamental performance metric is called the SIR meta distribution \cite{haenggi2015meta}, defined as a complementary cumulative distribution function (CCDF) of SIR success probability given the realization \cite{haenggi2021meta1,haenggi2021meta2}.

While SIR/SINR meta approximation is an important performance metric, it is difficult to compute since the exact integral expression is derived by using Gil-Pelaez theorem \cite{gil1951note} which requires the imaginary moments of the conditional success probability. Generally, beta approximation is applied to obtain the approximated SIR/SINR meta distribution, which only requires the first and the second moments of the conditional success probability. However, even deriving the first and the second moments of the conditional success probability is not always simple. Therefore, given the importance of SIR/SINR meta distribution and the difficulties of deriving it,  in this work, we provide an alternative approximation to  SIR/SINR meta which mainly requires distance distributions.

We propose a  novel approach to compute SINR meta distribution which relies on an approximation mix of exact and mean-field analysis of interference and reduces the complexity of the analysis and enhances tractability. In this paper, we call the proposed approximation 'dominant interferer-based approximation', however, we would like to clarify that this approximation is obtained by considering the dominant interferer(s) exactly while the rest are in the average sense. In other words, we considered all the interference.

\subsection{Related Work}
Literature related to this work can be categorized into: (i)  the concept of meta distribution and existing approximations and (ii) the applications of meta distribution in different system models. A brief discussion on related works in each of these categories is discussed in the following lines.

Stochastic geometry is a strong mathematical tool that enables characterizing the statistics of various large-scale wireless networks. In the analysis of wireless networks with randomly deployed nodes based on stochastic geometry, the PPP is the most widely used model. The authors in \cite{elsawy2016modeling,elsawy2013stochastic} presented a  tutorial on the fundamental concepts of the point process, modeling the interference in large-scale networks, and a comprehensive survey on single-tier, multi-tier, and cognitive cellular networks. Most of the existing studies focus on spatial averaging performance metrics. However, it yields limited information about the individual links.
 
 The concept of the meta distribution can be traced back to \cite{ganti2010correlation}, where the authors computed the distribution of the outage probability in a large-scale wireless network  instead of the spatial average. They obtained the outage distribution bound by calculating the moments. Similar works about individual links can be found in \cite{baccelli2006aloha,sousa1990optimum,andrews2011tractable,nigam2014coordinated,haenggi2013diversity,zhang2014stochastic}.  The definition and examples, Poisson bipolar networks and PPP networks, of meta distribution in wireless networks was provided in \cite{haenggi2021meta1}.  Since computing the meta distribution requires calculating the moments of conditional success probability, the exact equation is hard to derive.Therefore, statistical inequalities and approximations are extremely useful. The beta approximation is the most common one, which only requires the first two moments \cite{haenggi2015meta}.  In \cite{haenggi2021meta2}, the author provided a closed-form result for the SIR meta distribution by only considered the nearest interferer in Poission and Poisson bipolar networks.  Fourier-Jacobi expansion of moments to reconstruct meta distribution is used in \cite{8421026}. The separability of computing SIR meta distribution for any independent fading is analyzed in \cite{9072358}, their results show that the separable form behaves as a good approximation of the SIR meta distribution in Ginibre and triangular lattice networks when the SINR threshold is chosen large enough. The authors in \cite{kalamkar2019simple} study the asymptotics of the moments as the SINR threshold approaches $0$ for general networks. They provide the meta distribution of general networks by shifting the meta distribution of Poisson networks, and the shift gain is obtained by computing the ratio of the mean interference-to-signal ratios between the point process under consideration and the PPP. Based on the moments of the conditional success probability,  the authors in \cite{haenggi2018efficient,guruacharya2018approximation} provide the numerical methods to calculate the meta distribution.

 The meta distribution of both Poisson and non-Poisson cellular networks has already been addressed in few works. For instance, authors in  \cite{kouzayha2021meta} analyzed the binomial Poisson process (BPP)-based networks, and  \cite{saha2020meta,kalamkar2019simple} utilized the deployment gain to approximate the meta distribution for general networks. The SIR meta distribution of a 1-D hardcore point process was investigated in \cite{8830403}. Authors in \cite{9462466} studied a Poisson line Cox bipolar network and a Poisson stick line Cox bipolar network, and proposed a transdimensional PPP approximation, which highly reduces the complexity and improves the tractability of vehicular network analysis. SIR meta distribution for cellular networks with BS cooperation is analyzed in \cite{wang2017meta,wang2018sir,deng2019sinr,salehi2018meta,cui2017sir}, with power control, offloading, or with non-orthogonal multiple access (NOMA), respectively. SIR meta distribution of the moving networks is analyzed in \cite{9007512}. In \cite{mankar2019meta}, the authors characterized the meta distribution of the downlink SIR for the typical cell in the case of the BSs are modeled by PPP. The meta distribution of downlink SIR in a Poisson cluster process-based HetNet network was analyzed in \cite{saha2020meta}, in which the authors considered a $K$-tier HetNet modeled by a combination of PPP and PCP. Interestingly, they used mapping theorem \cite{madhusudhanan2016analysis,stoyan2013stochastic} to map the interference from $i$-th tiers onto one tier, which forms a new and unknown distribution, and obtained the meta distribution based on the new interference. 
 
 Unlike existing literature, this work aims to provide an alternative approximation to compute the meta distribution in the case of downlink Poisson  networks with Rayleigh fading without requiring computing the moments of success probability. Instead, the proposed approximation mainly requires computing the joint distance distribution of the first and the second nearest nodes in the network. 
\subsection{Contribution}
This paper derives a dominant interferer-based approximation of SINR meta distribution in downlink networks. Different from existing literature which mainly focused on beta distribution and approximated the moments of success probability, we consider a novel method that approximates the interference, consequently, the SINR meta distribution can be approximated by using a Lambert W function. The resulting approximation shows a good matching in different scenarios with both beta approximation and Monte-Carlo simulations in a large range of BS densities.

The contributions of this paper are:

\begin{itemize}
	\item We first derive approximate analytical expressions of the downlink conditional success probability for the PPP networks with a standard path loss model, and all the channels are subject to Rayleigh fading, which is based on the exact term of the dominant interferer(s) and the mean of the remaining interferers.
	
	\item We obtain the expression of the dominant interferer-based approximation of the SINR meta distribution, which can be written as a Lambert W function and mainly requires computing the joint CDF and the PDF of the first and the second nearest BS in the network. We show that the proposed approximation can be extended to the $j$-th nearest interfering BSs, which is more general, but the nearest interferer is already accurate enough, e.g., the gap between the exact value and the proposed approximation in the case of PPP networks with $\lambda = 1$ km$^{-2}$ and $\theta = 26$ dB is about $0.02$,   while at most be about $0.03$ at other points. We also show that the proposed approximation performs well at large path loss scenarios, say $\alpha = 3.5,4$.
	
	\item By applying the proposed approximation, we obtain the SINR meta distribution in four different wireless networks modeled by Poisson bipolar network, $K$-tier PPP, Mat\'{e}rn cluster process, and PLCP, and show that the proposed approximation is good matching and simple to compute since it mainly requires the joint distance distribution.
	
	\item Additionally, we derive the SINR meta distribution of the PLCP network , which has not been derived in existing literature, to the best of the authors' knowledge. Our results reveal that the proposed approximation highly reduces the complexity of computing its SINR meta distribution.
\end{itemize}

\section{The SINR Meta Distribution}
In this section, we introduce the SINR meta distribution based on a standard PPP wireless cellular network and focus on the downlink transmission in this network where the base stations (BSs) are modeled by a  PPP, denoted by $\Phi$, with density $\lambda$. We simply assume that all the BSs are active. In particular, we assume that the density of cellular users are much larger than the density of BSs, which leads to all BSs being active, and the user connects with the nearest BS since it provides the strongest average received power. We are interested in the SINR meta distribution of the network and its approximations.

Let $x_i$ be the locations of the BSs, where $x_i\in\Phi,i\in\mathbb{N}\cup \{0\}$ and $x_0$ is location of the nearest BS to the origin. Here, we condition a user to be at the origin and this user becomes the typical user on averaging over the point process. We focus on the SINR of the typical user which is equivalent to any other arbitrary deterministic location owing to the stationarity of PPP.
We use a standard path-loss model with exponent $\alpha>2$ and $h_{{x_i}}$ models the small scale Rayleigh fading of the channel between the typical user and the $i$-th BS, which is i.i.d and follows the exponential distribution with mean of unity.

Let $p_t$ be the transmit power of the BSs, the SINR at the typical user is
	\begin{align}
	{\rm SINR} = \frac{p_t h_{x_0} ||x_0||^{-\alpha}}{\sum_{x\in\Phi\setminus\{x_0\}}p_t h_{x} ||x||^{-\alpha}+\sigma^2},
\end{align}
where $\sigma^2$ is thermal noise. Consequently, the conditional success probability of the typical link is given by
\begin{align}
	P_s(\theta) = \mathbb{P}({\rm SINR}>\theta|\Phi).\label{eq_Ps}
\end{align}

For an arbitrary realization of $\Phi$, we analyze the fraction of links that exceed the reliability threshold $\theta$, and the target reliability is an argument to the SINR meta distribution. With that being said, our goal is to obtain the percentiles of users that achieve downlink coverage (SINR above $\theta$) at an arbitrary realization of the PPP network.
\begin{definition}[Meta Distribution]
	Generally, the SINR meta distribution of downlink is defined in \cite{haenggi2015meta} as 
	\begin{align}
		\bar{F}(\theta,\gamma) = \mathbb{P}(P_s(\theta) > \gamma),\label{eq_MetaF}
	\end{align}
	where $\gamma \in [0,1]$ and $P_s(\theta)$ is known as the success probability given the realizations of $\Phi$, i.e. the locations of BSs.
\end{definition}

\section{Mathematical Analysis}

Before investigating the SINR meta distribution, we first introduce some important distance distributions in PPP networks. In the following text, let $R_i = ||x_i||$ be the distance from the $i$-th interfering BS to the typical user and recall that $x_0$ is location of the serving BS. To simplify the notation,  we use $x_i$ to present the locations of BSs  which are ordered by the distances to the origin. That is,  $x_j$ presents the $j$-th nearest interfering BS  and $R_j$ denotes the distance to the $j$-th nearest interferer BS.
\begin{lemma}[Distance Distribution]\label{lemma_Dis}
	 As mentioned above, $R_0$ is the distance between the user and the nearest BS, which is the serving BSs, and $R_j$ is the distance to the $j$-th closest BS. The joint and marginal distance distributions are respectively given in \cite{moltchanov2012distance} as 
	\begin{align}
		f_{R_0,R_1,\cdots,R_j}(r_0,r_1,\cdots,r_j) &= (2\lambda\pi)^{j+1}r_{0}r_{1}\cdots r_{j}\exp(-\lambda\pi r_{j}^{2}),\nonumber\\
		f_{R_0,R_1}(r_0,r_1) &= (2\pi\lambda)^2r_0 r_1 \exp(-\pi\lambda r_1^2), \nonumber\\
		f_{ R_1}(r_1) &= 2(\pi\lambda)^2 r_1^3 \exp(-\pi\lambda r_1^2),\nonumber\\
		f_{R_0}(r_0) &= 2\pi\lambda r_0\exp(-\pi\lambda r_0^2),
	\end{align}
	consequently,  the  conditional PDF and CDF are
\begin{align}
		&f_{R_0\mid R_1}(r_0 \mid R_1) = \frac{2 r_0}{R_1^2}, \quad (0<r_0<R_1)\\
		&f_{R_1,\cdots,R_j}(r_1,\cdots,r_j) = \frac{f_{R_0,R_1,\cdots,R_j}(r_0,r_1,\cdots,r_j)}{f_{ R_0\mid R_1}(r_0 \mid R_1)},\label{eq_joint_r1rj}\\
		&F_{R_0\mid R_1}(r_0 \mid R_1) = \frac{r_{0}^{2}}{R_1^2}, \quad (0<r_0<r_1).\label{eq_conditionalCDFR0}
	\end{align}
\end{lemma}

After obtaining the distance distributions, we are able to compute the SINR meta distribution and its approximation, which are provided in the next two subsections.

\subsection{Existing Methods for Computing SINR Meta Distribution}

Clearly, SINR meta distribution is a two parameters complementary cumulative distribution function (CCDF).  We first give the exact expression of SINR meta distribution. Typically, (\ref{eq_MetaF}) is solved by using the Gil-Pelaez theorem \cite{haenggi2015meta,gil1951note}. The exact expression of SIR meta distribution is given in \cite{haenggi2015meta} and noise is added as an exponential factor to the moments.
\begin{theorem}	[Exact Expression of SINR Meta Distribution]\label{theorem_exactmeta}
	 The exact expression is given by
	\begin{align}
		\bar{F}(\theta,\gamma) &=  \frac{1}{2}+\frac{1}{\pi}\int_{0}^{\infty}\frac{\Im(\exp(-jt\log \gamma)M_{jt}(\theta))}{t}{\rm d}t,\label{eq_MetaExact}
	\end{align}
	where $j$ is the imaginary unit, $M_b(\theta)$ is the $b$-th moment of $P_s(\theta)$ and $\Im(\cdot)$ is the imaginary part of a complex number, and
\begin{align}
		&M_{b}(\theta) = \int_{0}^{\infty}\exp\bigg(-\lambda F_b \bigg)\exp\bigg(-\frac{b\theta}{p_t}r^{\alpha}\sigma^2\bigg)f_{R_0}(r){\rm d}r,
	\end{align}
	in which,
	\begin{align}
		F_b &= \pi\delta\sum_{k=1}^{\infty}\binom{b}{k}(-1)^{k+1} \frac{\alpha \theta^k R_0^{2}}{-2+k\alpha}{}_{2}F_{1}(k,-\delta+k,1-\delta+k,-\theta),\label{eq_Fb}
	\end{align}
where $\delta = 2/\alpha$ and ${}_{2}F_{1}$ is the Gaussian hypergeometric function.
\end{theorem}
\begin{IEEEproof}
See Appendix \ref{app_exactmeta}.
\end{IEEEproof}
However, (\ref{eq_MetaExact}) is difficult to compute since it requires the imaginary moments and three nested integrals. The beta distribution approximation is widely used to approximate the SINR meta distribution since their higher moments are very close \cite{haenggi2015meta}, and it only requires the first and the second moments of the success probability.
\begin{remark}[Beta Approximation]
	Using the beta approximation as mentioned in \cite{haenggi2015meta}, the  SINR meta distribution can be approximated as
	\begin{align}
		\bar{F}^{\prime}(\theta,\gamma)	\approx 1-I_{\gamma}\bigg(\frac{M_1(\theta)(M_1(\theta)-M_2(\theta))}{M_2(\theta)-M_1^2(\theta)},\frac{(M_1(\theta)-M_2(\theta))(1-M_1(\theta))}{M_2(\theta)-M_1^2(\theta)}\bigg),\label{eq_MetaBetaApp}
	\end{align}
	where,
	\begin{align}
		I_x(a,b) = \frac{\int_{0}^{x}t^{a-1}(1-t)^{b-1}{\rm d}t}{B(a,b)},
	\end{align}
	and $B(a,b) = \int_{0}^{1}t^{a-1}(1-t)^{b-1}{\rm d}t$.
\end{remark}
As shown in (\ref{eq_MetaBetaApp}), beta approximation only requires to compute the first two moments, which highly reduces the computing complexity.

\subsection{Proposed Approximation}

 Notice that the closer interference has the higher impact on the system performance compared to the rest of the interferers. Here, we provide an alternative approximation which approximates the interference by considering the $j$-th nearest interfering BSs and the conditional expectation of the sum of the remaining interfering BSs \cite{kishk2017joint,kishk2016downlink}.

We rewrite the aggregate interference based on the approximation policy mentioned above.

\begin{lemma}[Approximated Interference] 
	The interference at the typical user can be approximated as
\begin{align}
	I_j &\approx \sum_{k = 1}^{j} p_t h_{x_k} R_k^{-\alpha}+ p_t G(R_{j}),\label{eq_I}
\end{align}
where
\begin{align}
	G(R_{j}) = \frac{2\pi\lambda}{\alpha-2}R_{j}^{2-\alpha}. \label{eq_G}
\end{align}
\end{lemma}
\begin{IEEEproof}
	As mentioned, the closer interfering BS has higher impact on the system performance, hence, we consider the interference term composed of the exact expression of the closest $j$ interferers and the conditional mean of the rest of the terms,
	\begin{align}
		I_j &= \sum_{i\in\mathbb{N}}p_t h_{x_i} ||x_i||^{-\alpha}= \sum_{k = 1}^{j} p_t h_{x_k} R_k^{-\alpha}+ \sum_{k = j+1}^{\infty} p_t h_{x_k} R_k^{-\alpha}\nonumber\\
		&\approx \sum_{k = 1}^{j} p_t h_{x_k} R_k^{-\alpha}+p_t\mathbb{E}\bigg[\sum_{i\in\mathbb{N}\setminus\{1,\cdots,j\}} h_{x_i} ||x_i||^{-\alpha}\biggm\vert R_{j}\bigg],
	\end{align}
 and let $G(R_{j})$ denotes the average interference (without $p_t$) from  the remaining interferers, which is a function of the location of the $j$-th closest interferer, $R_j$, given by
\begin{align}
		G(R_{j}) &= \mathbb{E}\bigg[\sum_{i\in\mathbb{N}\setminus\{1,\cdots,j\}} h_{x_i} ||x_i||^{-\alpha}\biggm\vert R_{j}\bigg]\nonumber\\
		&\stackrel{(a)}{=} \mathbb{E}\bigg[\sum_{i\in\mathbb{N}\setminus\{1,\cdots,j\}} R_{i}^{-\alpha}\biggm\vert R_{j}\bigg]\stackrel{(b)}{=} \frac{2\pi\lambda}{\alpha-2}R_{j}^{2-\alpha},
	\end{align}
	where step $(a)$ follows from the assumption that all fading gains are independent and exponentially distributed with mean of unity and step $(b)$ follows Campbell's theorem \cite{haenggi2012stochastic} with conversion from Cartesian to polar coordinates.
\end{IEEEproof}


Approximated success probability is the final requirement to derive the approximated SINR meta distribution. Based on the approximation of the aggregate interference, we rewrite the interference term in SINR and the success probability is given in the following lemma.
\begin{lemma}[Approximated Conditional Success Probability]
	
	The conditional success probability is approximated by
	\begin{align}
		&P_{s,j}(\theta) \approx \exp\bigg(-\frac{\theta}{p_t}R_0^{\alpha}(p_t G(R_j)+\sigma^2)\bigg)\bigg(\frac{1}{1+\theta R_0^{\alpha}\sum_{k = 1}^{j} R_k^{-\alpha}}\bigg).\label{eq_approx_ps}
	\end{align}
\end{lemma}
\begin{IEEEproof}
Recall that	$R_i = ||x_i||$ is the distance from the $i$-th closest interfering BS to the typical user and $R_0$ is the distance to the serving BS. The conditional success probability is 	
\begin{align}
	\label{eq_appPs}
	P_{s,j}(\theta) &= \mathbb{P}\bigg(\frac{p_t h_{x_0} R_0^{-\alpha}}{\sum_{i\in\mathbb{N}}p_t h_{x_i} R_i^{-\alpha}+\sigma^2}>\theta\biggm\vert \Phi\bigg)\nonumber\\
	&= \mathbb{P}\bigg(h_{x_0}>\frac{\theta}{p_t}R_0^{\alpha}(\sum_{k = 1}^{j}p_t h_{x_k} R_k^{-\alpha}+\sum_{i\in\mathbb{N}\setminus\{1,\cdots,j\}}p_t h_{x_i} R_i^{-\alpha}+\sigma^2)\biggm\vert \Phi\bigg)\nonumber\\
	&\approx \mathbb{E}_{h_{\{x_k\}}}\bigg[\exp\bigg(-\theta R_0^{\alpha}\sum_{k = 1}^{j} h_{x_k} R_k^{-\alpha}-\theta R_0^{\alpha}G(R_j)-\frac{\theta}{p_t}R_0^{\alpha}\sigma^2\bigg)\bigg]\nonumber\\
	&= \exp\bigg(-\frac{\theta}{p_t}R_0^{\alpha}(p_t G(R_j)+\sigma^2)\bigg)\prod_{k = 1}^{j}\bigg(\frac{1}{1+\theta R_0^{\alpha} R_k^{-\alpha}}\bigg)\nonumber\\
	&\stackrel{(a)}{\approx}\exp\bigg(-\frac{\theta}{p_t}R_0^{\alpha}(p_t G(R_j)+\sigma^2)\bigg)\bigg(\frac{1}{1+\theta R_0^{\alpha}\sum_{k = 1}^{j} R_k^{-\alpha}}\bigg),
\end{align}
in which step (a) follows by ignoring the higher order terms of $\theta(\frac{R_0}{R_j})^{\alpha}$.	
\end{IEEEproof}
Now we are able to proceed to the final expression of  the proposed  approximation. 
Since there is less multiplication in conditional success probability in (\ref{eq_approx_ps}) compared to (\ref{eq_ps}), the SINR meta distribution can be easily obtained.
\begin{theorem}[Approximated Meta Distribution]
		The approximated SINR meta distribution, $\bar{F}_{j}^{\prime}(\theta,\gamma)$, is given by
	\begin{align}
		&\bar{F}_{j}^{\prime}(\theta,\gamma) \approx\int_{(\mathbb{R}^{+})^j}F_{R_0|R_1}(K_j(r_1,r_2,...,r_j))f_{ R_1,R_2,...R_j}(r_1,r_2,...,r_j){\rm d}r_1{\rm d}r_2,...{\rm d}r_j,\label{eq_j_appMeta}
	\end{align}
which is a $j$-dimensional integral, and
	\begin{align}
		&K_j(r_1,r_2,...,r_j) =\bigg(-\frac{1}{\theta \sum_{k = 1}^{j}r_k^{-\alpha}}+\frac{1}{s(r_j)}W\bigg(0,\frac{s(r_j)\exp(s(r_j)\theta^{-1}(\sum_{k = 1}^{j}r_k^{-\alpha})^{-1})}{\gamma\theta r_1^{-\alpha}}\bigg)\bigg)^{\frac{1}{\alpha}},
	\end{align}
	where $W(k,x)$ is the $k$-th branch of the Lambert W function, $s(r_j) = \frac{\theta}{p_t}(p_t G(r_j)+\sigma^2)$, $F_{ R_0|R_1}(r)$ and $f_{R_1,R_2,...R_j}(r_1,r_2,...,r_j)$ are given in (\ref{eq_joint_r1rj}) and (\ref{eq_conditionalCDFR0}).
\end{theorem}
\begin{IEEEproof}
	By using the definition of SINR meta distribution, the CCDF of the conditional success probability is given by
	\begin{align}
		\label{eq_poof_MD}
		&\bar{F}_{j}^{\prime}(\theta,\gamma) = \mathbb{P}(P_{s,j}(\theta)>\gamma)\nonumber\\
		&= \mathbb{E}_{ R_1,\cdots,R_j}\bigg[\mathbb{P}\bigg(\exp\bigg(-\frac{\theta}{p_t}R_0^{\alpha}(p_t G(R_j)+\sigma^2)\bigg)\prod_{k = 1}^{j}\bigg(\frac{1}{1+\theta R_0^{\alpha} R_k^{-\alpha}}\bigg)>\gamma\mid  R_1,\cdots,R_j\bigg)\bigg] \nonumber\\
		&= \mathbb{E}_{ R_1,\cdots,R_j}\bigg[\mathbb{P}\bigg(\exp(-s(R_j) R_0^{\alpha})>\gamma\prod_{k = 1}^{j}\bigg(1+\theta R_0^{\alpha} R_k^{-\alpha}\bigg) \mid R_1,\cdots,R_j\bigg)\bigg] \nonumber\\
		&\approx \mathbb{E}_{ R_1,\cdots,R_j}\bigg[\mathbb{P}\bigg(\exp(-s(R_j) R_0^{\alpha})>\gamma\bigg(1+\theta R_0^{\alpha}\sum_{k = 0}^{j} R_k^{-\alpha}\bigg) \mid  R_1,\cdots,R_j\bigg)\bigg] \nonumber\\
		&\stackrel{(a)}{=} \mathbb{E}_{ R_1,\cdots,R_j}\bigg[\mathbb{P}\bigg(R_0<\bigg(-\frac{1}{\theta \sum_{k = 1}^{j}R_k^{-\alpha}}+\nonumber\\
		&\qquad\qquad\quad\frac{1}{s(R_j)}W\bigg(0,\frac{s(R_j)\exp(s(R_j)\theta^{-1}(\sum_{k = 1}^{j}R_k^{-\alpha})^{-1})}{\gamma\theta \sum_{k = 1}^{j}R_k^{-\alpha}}\bigg)\bigg)^{\frac{1}{\alpha}}\mid R_1,\cdots,R_j\bigg)\bigg],
	\end{align}
	where step $(a)$ follows from that Lambert W function defined by: $W(0,x)\exp(W(0,x)) = x$,  and this step can be solved by using {\rm MATLAB} or {\rm Mathematica}, and the proof completes by using the conditional CDF of $R_0$ and the joint PDF of $R_1,\cdots,R_j$ provided in Lemma \ref{lemma_Dis}.
\end{IEEEproof}

	Observing that (\ref{eq_j_appMeta}) requires to computing the joint PDF from $R_1$ to $R_j$, we observe that it is still too complex and not practical in general. Considering that the nearest interferer BS has the highest impact on the system performance, we further simplify the proposed approximation which only composed of the exact expression of the nearest interferer and the conditional mean of the rest of the terms.

We would like to clarify that the approximation proposed in Cor. 1 is considering the dominant interferer exactly while the rest are considered in an average sense. However, we name it the dominant interferer-based approximation to simplify the name.

\begin{cor}[The Dominant Interferer-based Meta distribution] 
	\label{cor_dom}
	 If we only consider the first nearest interferer BS exactly while the rest are considered in an average sense, (\ref{eq_j_appMeta}) can be further simplified,
\begin{align}
\label{eq_MetaOurApp}
&\bar{F}^{\prime}_{1}(\theta,\gamma) \approx\int_{0}^{\infty}F_{R_0|R_1}(K_1(r,\theta,\gamma))f_{R_1}(r){\rm d}r,
\end{align}
where,
\begin{align}
	K_1(r,\theta,\gamma) = \bigg(-\frac{1}{\theta r^{-\alpha}}+\frac{1}{s(r)}W\bigg(0,\frac{s(r)\exp(s(r)\theta^{-1}r^{\alpha})}{\gamma\theta r^{-\alpha}}\bigg)\bigg)^{\frac{1}{\alpha}},\label{eq_h1}
\end{align}
in which $s(r) = \frac{\theta}{p_t}(p_t G(r)+\sigma^2)$ and $G(r) = \frac{2\pi\lambda}{\alpha-2}r^{2-\alpha}$.
\end{cor}
\begin{IEEEproof} 
	Proof completes by setting $j = 1$ in (\ref{eq_j_appMeta}).
\end{IEEEproof}

\begin{remark}
	\label{rem_nearest}
In the case of $\sigma^2 = 0$, e.g., approximation of SIR meta distribution, $s(r) = \theta G(r)$, (\ref{eq_h1}) can be further simplified,
\begin{align}
	K_{1}^{\prime}(r,\theta,\gamma) = \bigg(-\frac{1}{\theta r^{-\alpha}}+\frac{1}{s(r)}W\bigg(0,\frac{s^{\prime}(r)}{\gamma}\exp(s^{\prime}(r))\bigg)\bigg)^{\frac{1}{\alpha}},\label{eq_h1_pri}
\end{align}
where $s^{\prime}(r) = \frac{2\pi\lambda}{(\alpha-2)}r^{2}$. Moreover, if we only consider the dominant interferer, that is $G(r) = 0$,
\begin{align}
	K_{1}^{\prime\prime}(r,\theta,\gamma) = \bigg(\frac{1}{\gamma\theta r^{-\alpha}}-\frac{1}{\theta r^{-\alpha}}\bigg)^{\frac{1}{\alpha}} = r\bigg(\frac{1-\gamma}{\gamma\theta}\bigg)^{\frac{1}{\alpha}},\label{eq_h1_pri_p}
\end{align}
consequently, (\ref{eq_MetaOurApp}) becomes
\begin{align}
\bar{F}^{\prime\prime}(\theta,\gamma) = \min\bigg(1,\bigg(\frac{1-\gamma}{\gamma\theta}\bigg)^{\frac{1}{\alpha}}\bigg),
\end{align}
which is the same as the result of the nearest-interferer-only approximation mentioned in \cite[Cor.~3]{haenggi2021meta2}.  

The difference and relation between the proposed approximation, dominant-interferer approximation (\ref{eq_MetaOurApp}), and the nearest-interferer-only approximation, \cite[Cor.~3]{haenggi2021meta2}, are: (i) the nearest-interferer-only approximation is an upper bound of the meta distribution while the proposed approximation is neither upper bound nor lower bound and it is tighter (more details shown in Numerical Results Section), (ii) the proposed approximation considered all the interferer while the result in \cite[Cor.~3]{haenggi2021meta2} removed some interferers, and (iii) the proposed approximated SINR meta distribution is an approximation of exact SINR meta distribution and nearest-interferer-only meta distribution and SIR meta distribution are special cases of the proposed approximation.
\end{remark}
The SINR meta distribution is computed by the joint CDF and the PDF of the first and the second nearest BSs, which are the fundamental distance distributions of point processes. Hence, the proposed approximation has different requirements   compared with the beta approximation. In the point processes which are able to obtain the joint distance distributions, the proposed approximation can be a candidate approach to approximate the SINR meta distribution. In the numerical results part, we show that the dominant interferer-based approximation is already very accurate, e.g., the gap between the exact value and the proposed approximation in the case of PPP networks with $\lambda  = 1$ km$^{-2}$ and $\theta = 26$ dB is about $0.02$, while the gaps for the remaining points are at most about $0.03$, in the downlink SINR meta distribution of Poisson network with Rayleigh fading.

\section{The Proposed Approximation in Poisson Networks}
In this section, we use the proposed approximation in three wireless cellular networks modeled by three commonly used point processes, respectively, and one ad hoc network, to show its accuracy and tractability. The one ad hoc scenario and  three cellular scenarios are (i) Poisson bipolar networks, (ii) Mat\'{e}rn cluster process (MCP), (iii) $K$-tier Poisson network, and (iv) Poisson line Cox process (PLCP), respectively. In (i), (ii) and (iii), we give the expressions of the SINR meta distribution for both beta approximation and the proposed approximation. In (iv), we use the proposed approximation under PLCP-distributed interference and PPP-approximated interference, respectively, and compared the proposed approximation with a transdimensional PPP-based (TPPP-based) approximation as mentioned in \cite[(16)]{9462466}.

\subsection{Poisson Bipolar Networks}
In this part, we consider a Poisson bipolar model, where the transmitters form a PPP, denoted by $\Phi$ with density $\lambda$, and each of the transmitter has a dedicated receiver at distance $R$ in a random direction. Let $p_t$ be the transmit power and all channels are subject to Rayleigh fading, and we use the standard path loss model with exponent $\alpha$.

Before investigating the SINR meta distribution, we first introduce some important distance distributions in Poisson bipolar networks. Note that in Poisson bipolar network, each receiver has a dedicate transmitter at distance $R$. Therefore, the distance to the serving BS is determined and independent from the nearest interferer, while the distance distribution of the nearest interferer follows the first contact distance distribution in PPP.
The CDF of $R_0$ given $R_1$ equals to the CDF of $R_0$, which is a shifted step function, and   $R_1$ is the the contact distance in PPP,
\begin{align}\label{eq_distance_PB}
	f_{ R_0}(r) &=  \Delta_{R}(r), \nonumber\\
	F_{R_0}(r) &=  \left\{ 
	\begin{aligned}
		0,  & \quad  r < R,\\
		1,  & \quad r\geq R,\\
	\end{aligned} \right.,\nonumber\\
f_{ R_1}(r ) &= 2\pi\lambda r \exp(-\pi\lambda r^2),
\end{align}	
in which $\Delta_{a}(x)$ is an impulse at $a$ and satisfies $\int_{0}^{\infty} \Delta_{a}(r)dr=1$.

 The SIR success probability and $b$-th moment of  the conditional success probability of a Poisson bipolar network are well known \cite{haenggi2012stochastic,haenggi2015meta} and given by
\begin{align}
	P_{s,{\rm SIR}}(\theta) &= \exp(-C\theta^{\delta}),\nonumber\\
	M_{b,{\rm SIR}}(\theta) &= \exp\bigg(-C\theta^{\delta}\frac{\Gamma(b+\delta)}{\Gamma(1+\delta)\Gamma(b)}\bigg),
\end{align}
where $\delta = 2/\alpha$ and $C$ is a coefficient that does not depend on $\theta$: $C = \lambda\pi R^2\Gamma(1-\delta)\Gamma(1+\delta)$.
Therefore, the SINR success probability and the $b$-th moment are obtained by adding the exponential term and given by, respectively,
\begin{align}
	P_s(\theta) =& P_{s,{\rm SIR}}(\theta)\exp\bigg(-\frac{\theta R^{\alpha}}{{p_t}}\sigma^2\bigg),\nonumber\\
	M_b(\theta)	=& \exp\bigg(-C\theta^{\delta}\frac{\Gamma(b+\delta)}{\Gamma(1+\delta)\Gamma(b)}\bigg)\exp\bigg(-b\frac{\theta R^{\alpha}}{{p_t}}\sigma^2\bigg).\label{eq_PB_Mb}
\end{align}

Therefore, the beta approximation of the SINR meta distribution of a Poisson bipolar network is obtained by substituting (\ref{eq_PB_Mb}) into (\ref{eq_MetaBetaApp}), and the proposed approximation of the SINR meta distribution  is obtained by substituting (\ref{eq_distance_PB}) into (\ref{eq_MetaOurApp}).
\subsection{Mat\'{e}rn Cluster Process}

In this part, we consider a Mat\'{e}rn cluster point process to model the locations of users and BSs \cite{afshang2017nearest}. In MCP, the clusters are modeled as disks with radii $r_c$ whose centers are modeled as a PPP, $\Phi$ with density $\lambda$, while the users are uniformly distributed within the disk. Assume that the user associates with its cluster BS. Let $p_t$ be the transmit power of BSs and all channels are subject to Rayleigh fading, and we use the standard path loss model with exponent $\alpha$.

In order to use the proposed approximation, some distance distributions are given below. Since we assume that the user associates with its cluster BS, the distance to the serving BS is independent from the nearest interferer.
The CDF of $R_0$ given $R_1$ equals to the CDF of $R_0$ and   $R_1$ is the the first contact distance in PPP,
\begin{align}
	f_{ R_0}(r) &=  \frac{2r}{r_{c}^{2}},  \quad  r < r_c,\nonumber\\
	f_{ R_1}(r ) &= 2\pi\lambda r \exp(-\pi\lambda r^2),\nonumber\\
	F_{ R_0}(r) &=  \frac{r^2}{r_{c}^{2}},  \quad  r < r_c,\label{eq_distance_MCP}
\end{align}
where $r_c$ is the user cluster radius.

	The SINR success probability and the $b$-th moment are, respectively, given by
\begin{align}
	P_s(\theta) &= \mathbb{P}({\rm SINR} > \theta) = \mathbb{P}\bigg(\frac{p_t h R_{0}^{-\alpha}}{I+\sigma^2} > \theta\bigg) = \mathbb{P}\bigg( h > \frac{\theta R_{0}^{\alpha}}{p_t}(I+\sigma^2)\bigg) = \exp\bigg(-g(R_{0})(I+\sigma^2)\bigg)\nonumber\\
	&= \prod_{i\in\mathbb{N}}\bigg(\frac{1}{1+g(R_{0})p_t R_i^{-\alpha}}\bigg)\exp(-g(R_{0})\sigma^2),\nonumber\\
	M_b(\theta) &= \mathbb{E}\bigg[\prod_{i\in\mathbb{N}}\bigg(\frac{1}{1+g(R_{0})p_t R_i^{-\alpha}}\bigg)^b\exp(-bg(R_{0})\sigma^2)\bigg]\nonumber\\
	&= \int_{0}^{r_c}\exp\bigg(-2\pi\lambda\int_{0}^{\infty}\bigg[1-\bigg(\frac{1}{1+g(r)p_t z^{-\alpha}}\bigg)^b\bigg]z{\rm d}z\bigg)\exp(-bg(r)\sigma^2)f_{R_0}(r){\rm d}r,\label{eq_MCP_Mb}
\end{align}
where $g(r) = \frac{\theta r^{\alpha}}{p_t}$.

Therefore, the beta approximation of the SINR meta distribution of a MCP network is obtained by substituting (\ref{eq_MCP_Mb}) into (\ref{eq_MetaBetaApp}), and the proposed approximation of the SINR meta distribution  is obtained by substituting (\ref{eq_distance_MCP}) into (\ref{eq_MetaOurApp}).

\subsection{$K$-tier Poisson Point Process}
In this part, we consider a general $K$-tier cellular network model, where BSs of each tier follow a homogeneous independent PPP $\Phi_i$ with density $\lambda_{i}$. For the BSs in the $i$-th tier, the transmit power is $p_{t,i}$. Assume that the user associates with the BS that provides   the strongest average received power.

In order to use the proposed approximation, which is a function of the first and the second nearest BSs, we map the BSs on the $i$-th tier to the $1$-st tier to obtain the distance distributions of the first and the second nearest BSs. After mapping, the equivalent density in the $1$-st tier is
\begin{align}
	\lambda_{i}^{\prime} = (\frac{p_{t,1}}{p_{t,i}})^{-\delta}\lambda_{i}.
\end{align}
By the results of linear mapping and the superposition property of the point process, the new $1$-tier network is still a homogeneous PPP $\Phi^{\prime}$ with density
\begin{align}
	\lambda^{\prime} = \sum_{i = 1} (\frac{p_{t,1}}{p_{t,i}})^{-\delta}\lambda_{i}^{\prime}.\label{eq_equivalentdensity}
\end{align}

We then use the result from the single tier PPP network. The $b$-th moment is 
\begin{align}
	M_{b}(\theta) = \mathbb{E}\bigg[\exp\bigg(-\lambda^{\prime} F_b \bigg)\exp\bigg(-\frac{b\theta}{p_t}R_0^{\alpha}\sigma^2\bigg)\bigg],\label{eq_KPPP_Mb}
\end{align}
where $F_b$ is given in (\ref{eq_Fb}).

Therefore, the beta approximation of the SINR meta distribution of a $K$-tier PPP network is obtained by substituting (\ref{eq_KPPP_Mb}) into (\ref{eq_MetaBetaApp}), and the proposed approximation of the SINR meta distribution  is obtained by substituting the equivalent BS density in (\ref{eq_equivalentdensity}) and the distance distribution in Lemma \ref{lemma_Dis} into (\ref{eq_MetaOurApp}).

\subsection{Poisson Line Cox Process}
\begin{figure}[ht]
	\centering
	\includegraphics[width  = 0.8\columnwidth]{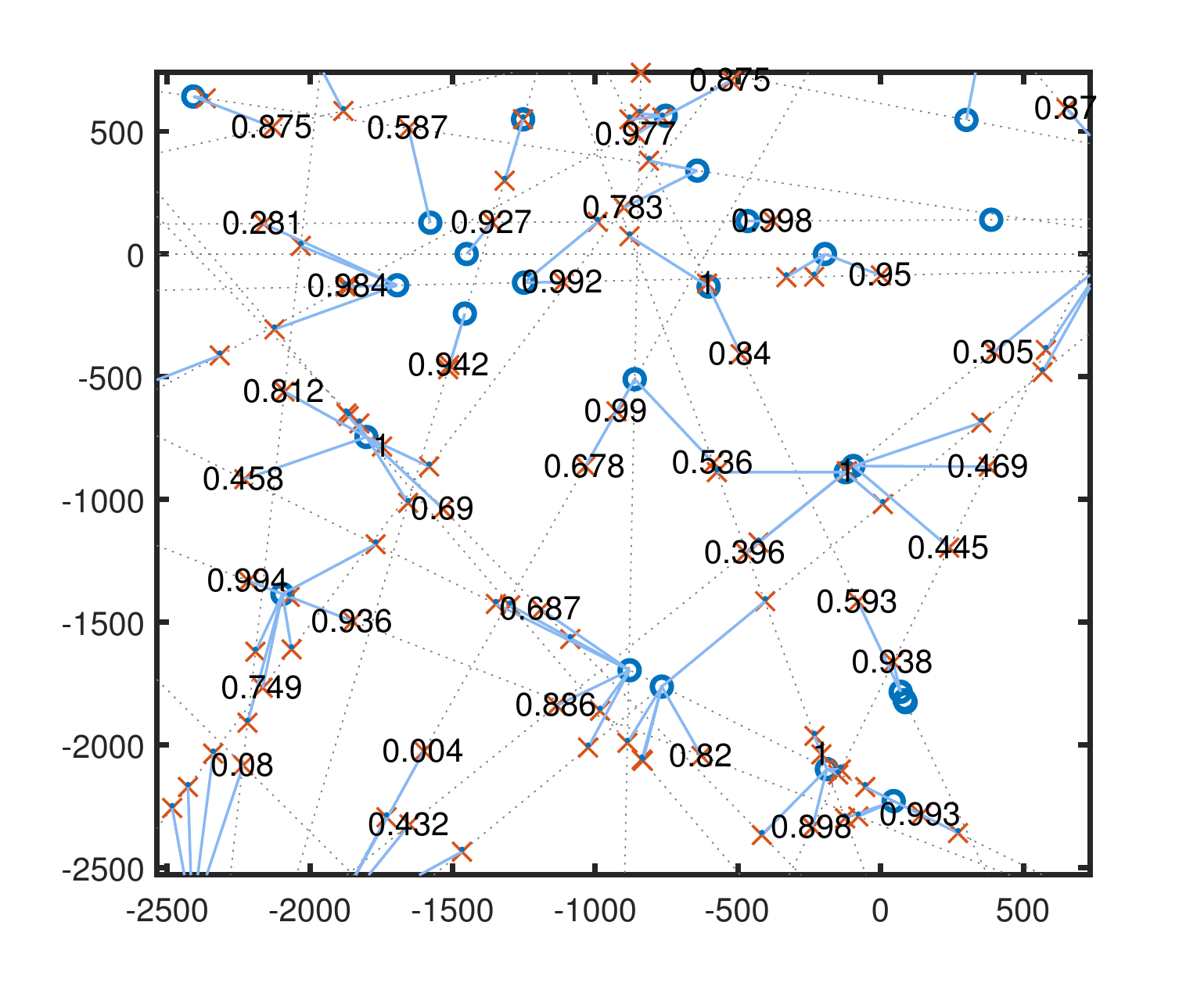}
	\caption{Illustration of the conditional SINR success probability of a Poisson line Cox process for $\lambda = 1.6$ BS/km$^2$, where $\lambda_l = 0.4/\pi$ km/km$^2$ and $\lambda_p = 0.4$ /km. The markers denote the BSs, users are indicated by crosses, blue segments denote the links and gray lines denote the Poisson line process. The number next to each link is its success probability.  Note that all the links have a number denoting the success probability, but due to the limitation of the space, some numbers may not show.}
	\label{fig_PLCP_sys}
\end{figure}
In this section, we consider a general Poisson line Cox process (PLCP) cellular network model, as shown in Fig. \ref{fig_PLCP_sys}, where the locations of BSs and users are modeled by the same Poisson line process, with line density $\mu_l = \pi\lambda_l$, where $\lambda_l$ is the point density of the corresponding point process in representation space, and two independent 1D PPPs, with densities $\lambda_{p}$ and $\lambda_{p,u}$, respectively. Let $\Phi$ be the point set of the locations of BSs and the corresponding density is $\lambda= \lambda_p\lambda_l\pi$. Assume that the user associates with the nearest BS. Noting that in PLCP network, we define the user point process since it  influence the distance distribution, e.g., a typical line passes though the user.

To obtain the distance distribution, we need to compute the probability of the number of the points falling in a unit region. In PLCP, the PMF of the number of points in $B(0,r)$ is given in \cite{choi2018poisson,dhillon2020poisson} and obtained by 
\begin{align}
	\mathbb{P}(\mathcal{N}_p(B(0,r)\cap\Phi_{0}^{!})=k)= (-\lambda_p)^k \frac{\partial^k\mathcal{L}^{!}_{(0,r)}(\lambda_p)}{\partial\lambda_{p}^{k}},	
\end{align}
where $\mathcal{N}_p(A\cap\Phi)$ denotes the counting measure, which counts the number of points in point set $\Phi$ falling in the A, $\Phi_{0}^{!}$ denotes the reduced Palm distribution \cite[Cor. 4.2]{dhillon2020poisson}, in which a typical line passes though the origin, $B(0,r)$ denotes the ball centered at the origin with radius $r$ and $\mathcal{L}^{!}_{(0,r)}(s)$ is Laplace transform of the total chord length in $B(0,r)$ under the Palm distribution,
\begin{align}
	\mathcal{L}_{(0,r)}^{!}(s) &= \exp \left[-2s r-2 \pi \lambda_l \int_{0}^{r} 1-\exp \left(-2 s \sqrt{r^{2}-\rho^{2}}\right) \mathrm{d} \rho\right].
\end{align}

\begin{lemma}[Distance Distributions of PLCP]
The PDF of $R_0$ and the PDF of $R_1$ are, respectively, given by
\begin{align}
f_{R_0}(r_0) &= 2\bigg(\lambda_p+\pi\lambda_l\int_{0}^{r_0}\frac{2r_0\lambda_p}{\sqrt{r_0^2-x^2}}\exp(-2\lambda_p\sqrt{r_0^2-x^2}){\rm d}x\bigg)\nonumber\\
&\quad\times\exp \left[-2\lambda_p r_0-2 \pi \lambda_l \int_{0}^{r_0} 1-\exp \left(-2 \lambda_p \sqrt{r_0^{2}-\rho^{2}}\right) \mathrm{d} \rho\right],\\
f_{\rm R_1}(r_1) 
& = \exp \left[-2\lambda_p r_1-2 \pi \lambda_l \int_{0}^{r_1} 1-\exp \left(-2 \lambda_p \sqrt{r_1^{2}-\rho^{2}}\right) \mathrm{d} \rho\right]\nonumber\\
&\quad\times \bigg[\lambda_p\bigg(-2+4r_1\lambda_p+4\lambda_p\lambda_l\pi\int_{0}^{r_1}2\sqrt{r_1^2-x^2}\exp(-2\lambda_p\sqrt{r_1^2-x^2}){\rm d}x\nonumber\\
&\qquad-2\pi\lambda_l\int_{0}^{r_1}\frac{r_1(2-4\lambda_p\sqrt{r_1^2-x^2})}{\sqrt{r_1^2-x^2}}\exp(-2\lambda_p\sqrt{r_1^2-x^2}){\rm d}x\nonumber\\
&\qquad+4\pi\lambda_l\int_{0}^{r_1}\frac{r_1\lambda_p}{\sqrt{r_1^2-x^2}}\exp(-2\lambda_p\sqrt{r_1^2-x^2}){\rm d}x\nonumber\\
&\quad\qquad\times(r_1+\pi\lambda_l\int_{0}^{r_1}2\sqrt{r_1^2-x^2}\exp(-2\lambda_p\sqrt{r_1^2-x^2}){\rm d}x)\bigg)\nonumber\\
&\quad+2(\lambda_p+\pi\lambda_l\int_{0}^{r_1}\frac{2r_1\lambda_p}{\sqrt{r_1^2-x^2}}\exp(-2\lambda_p\sqrt{r_0^2-x^2}){\rm d}x)\bigg].
\end{align}
\end{lemma}
\begin{IEEEproof}
The first contact distance distribution is derived by computing the void probability, which is defined as the probability of no points in $B(0,r)$. From the definition of PLCP, the void probability, which is also the CCDF of $R_0$, immediately follows,
\begin{align}
	\mathbb{P}(R_0>r_0) &= \mathbb{P}(\mathcal{N}_p(B(0,r_0)\cap\Phi_{0}^{!})=0)= \mathcal{L}^{!}_{(0,r_0)}(\lambda_p)\nonumber\\
	& =\exp \left[-2\lambda_p r_0-2 \pi \lambda_l \int_{0}^{r_0} 1-\exp \left(-2 \lambda_p \sqrt{r_0^{2}-\rho^{2}}\right) \mathrm{d} \rho\right],
\end{align}
$f_{\rm R_0}(r_0)$ then obtain by one minus the derivative of above equation over $r_0$. Similarly, the CCDF of the distance to the second neighbor in PLCP is derived by
\begin{align}
	\mathbb{P}(R_1>r_1) &= \mathbb{P}(\mathcal{N}_p(B(0,r_1)\cap\Phi_{0}^{!})=0)+\mathbb{P}(\mathcal{N}_p(B(0,r_1)\cap\Phi_{0}^{!})=1)\nonumber\\
	& = \mathcal{L}^{!}_{(0,r_1)}(\lambda_p)+(-\lambda_p)\frac{\partial \mathcal{L}^{!}_{(0,r_1)}(\lambda_p)}{\partial\lambda_p},
\end{align}
proof completes by taking the derivative over $r_1$,
\begin{align}
	f_{R_1}(r_1) &= -\frac{\partial \mathcal{L}_{(0,r_1)}^{!}(\lambda_p)}{\partial r_1}-(-\lambda_p)\frac{\partial^2 \mathcal{L}_{(0,r_1)}^{!}(\lambda_p)}{\partial\lambda_p\partial r_1}.\label{eq_Dis_PCLP}
\end{align}
\end{IEEEproof}

In what follows, we compute the conditional distance distribution in PLCP networks.

	\begin{lemma}[Conditional Distance Distribution of PLCP]
	The conditional CDF $F_{R_0|R_1}(r_0|R_1)$ is the final requirement to approach the proposed approximation of the SINR meta distribution,
\begin{align}
	F_{R_0|R_1}(r_0|R_1) =& \frac{f_{ R_0}(r_0)}{f_{ R_1}(R_1)}-\frac{f_{ R_0}(r_0)}{f_{ R_1}(R_1)} \exp(-2\lambda_p (R_1-r_0))\nonumber\\
	&\times\exp\bigg(-2\pi\lambda_l\int_{0}^{R_1}1-\exp\bigg(-2\lambda_p\bigg(\sqrt{R_1^2-\rho^2}-\sqrt{\max(0,r_0^2-\rho^2)}\bigg)\bigg){\rm d}\rho\bigg).\label{eq_conditionDis_PLCP}
\end{align}
\end{lemma}
\begin{IEEEproof}
We first obtain $F_{R_1|R_0}(r_1|r_0)$, which denotes the CDF of the distance to the second nearest neighbor (which is the probability that the distance to the second nearest neighbor is less than $r_1$) given the nearest one located at distance $r_0$.
To do so, we start with computing the CCDF,
\begin{align}
\bar{F}_{R_1|R_0}(r_1|R_0) &= \mathbb{P}(R_1 > r_1 | R_0)\nonumber\\
&= \mathbb{P}(\mathcal{N}_p(B(0,r_1)\cap\Phi_{0}^{!}-B(0,R_0)\cap\Phi_{0}^{!})=0) = \mathcal{L}_{(0,r_1)-(0,R_0)}^{!}(\lambda_p)\nonumber\\
&= \exp\bigg(-2\pi\lambda_l\int_{0}^{r_1}1-\exp\bigg(-2\lambda_p\bigg(\sqrt{r_1^2-\rho^2}-\sqrt{\max(0,r_0^2-\rho^2)}\bigg)\bigg){\rm d}\rho\bigg)\nonumber\\
&\quad\times\exp(-2\lambda_p (r_1-R_0)),
\end{align}
in which the subscript $(0,r_1)-(0,R_0)$ denotes the area $B(o,r_1)\setminus B(o,R_0)$, then the proof completes by applying the Bayes' theorem.
\end{IEEEproof}

Note that the interference distribution of PLCP networks is slightly different from PPP networks. Hence, the approximation of the interference term $G(R_1)$ in (\ref{eq_MetaOurApp}) should be recomputed, which is shown in the following lemma. 
\begin{lemma}[Interference Approximation of PLCP]
In PLCP, the conditional mean of the rest interference (without $p_t$), is given by
\begin{align} 
	\label{eq_G1}
	G_1(R_{1}) =& 2\pi\lambda_{l}\lambda_{p}\bigg(\int_{R_1}^{\infty}\int_{0}^{\infty}(\rho^2+r^2)^{\frac{-\alpha}{2}}{\rm d}r{\rm d}\rho+\int_{0}^{R_1}\int_{\sqrt{\rho^2-r^2}}^{\infty}(\rho^2+r^2)^{\frac{-\alpha}{2}}{\rm d}r{\rm d}\rho\bigg)\nonumber\\
	&+2\lambda_p\frac{R_{1}^{1-\alpha}}{\alpha-1}.
\end{align}
\end{lemma}
\begin{IEEEproof}
In PLCP, the remaining interference is composed of three parts: (i) interference from the typical line (the line passes though the origin, $L_0$), (ii) interference from the remaining lines, which do not intersect with $B(0,R_1)$ and (iii) interference from the remaining lines, which intersect with $B(0,R_1)$:
\begin{align}
&	G_1(R_{1}) = \mathbb{E}\bigg[\sum_{i\in\mathbb{N}\setminus\{1\}} h_{x_i} ||x_i||^{-\alpha}|R_1\bigg]\nonumber\\
	&= \mathbb{E}\bigg[\sum_{x\in L_0\setminus\{B(0,R_1)\}} ||x||^{-\alpha}|R_1\bigg]+\mathbb{E}\bigg[\sum_{x\in \Phi_0\setminus\{B(0,R_1)\},\rho<R_1} ||x||^{-\alpha}|R_1\bigg]+\mathbb{E}\bigg[\sum_{x\in \Phi_0\setminus\{B(0,R_1)\},\rho>R_1} ||x||^{-\alpha}|R_1\bigg],
\end{align} 
in which $\Phi_0 = \Phi_{0}^{!}\setminus L_0$, $\rho$ denotes the distance from the line to the origin, and the proof completes by using Campbell's theorem \cite{haenggi2012stochastic} with conversion from Cartesian to polar coordinates. 
\end{IEEEproof}
Since PPP is a good approximation of PLCP \cite{dhillon2020poisson}, we compute the proposed approximation based on (\ref{eq_G}) and (\ref{eq_G1}), respectively, in the numerical result section and compare these two approximations. 

Consequently,  the proposed approximation of the SINR meta distribution  is obtained by substituting the distance distribution in (\ref{eq_conditionDis_PLCP}) and (\ref{eq_Dis_PCLP}) into (\ref{eq_MetaOurApp}). Besides, authors in \cite{9462466} provides an interesting approximation to the SIR meta distribution of PLP with fixed transmission distance and we compare the proposed approximation to their approximation (by adding the noise and taking the expectation over $R_0$). 

\medskip

Here, we provide some advantages and limitations of the proposed approximation. The proposed approximation has different requirements, which are joint distance distributions and a spatial average of the interference except for the nearest interferer, compared to the traditional approximation. Therefore,  for the point processes where the joint distance distributions are available or can be approximated, the proposed approximation is simple since it only needs one integral which is about the distance to the nearest interferer (the expectation over $R_1$). Besides, computing moments is not always trivial, it depends on the system model, e.g., the point process. The moments of PPP, MCP, and Poisson bipolar networks are trivial, for PLCP, however, the moments are not straightforward owing to the locations of interferers. For instance in \cite[(7.46)]{dhillon2020poisson} is the Laplace transform of the interference in PLCP networks, which is the PGFL part of the moments. Therefore, the proposed dominant interferer-based approximation does provide some simplicity in computing SINR meta distribution. However, the proposed approximation is restricted by the ability to solve $R_0$ (step (a) in (\ref{eq_poof_MD})), e.g., it can not be solved under Nakagami-m fading, and for the point processes which has an unknown joint distance distribution, the proposed approximation may not work.

\section{Numerical Results}

In this section, we validate  the proposed  approximation via Monte-Carlo simulations with a large number of iterations ($5\times10^5$ iterations, in which $100$ for location realizations and $5000$ for fading realizations) to ensure the accuracy, and compare it with the beta approximation and the exact value of the SINR meta distribution. We start with the standard PPP network. We first generate two realizations of PPPs for the locations of users and BSs. Users are located in the Voronoi cells formed by BSs. While the realizations of PPPs are fixed, the fading realizations change in each iteration and we compute the success probability for each link. We then obtain the CCDF of the success probability of each realization. Unless stated otherwise, we use the system parameters listed herein. The transmit power of BS is $p_t = 10$ W, path-loss exponent is $\alpha = 4$, noise power is $\sigma^2 = 10^{-9}$ W and the density of BS changes from $0.1$ to $10$ BS/km$^{2}$. In the single-tier case  either $p_t$ or $\sigma^2$ can be set to $1$ since only their ratio matters, however,  we set the values of $p_t$ or $\sigma^2$ separately since we have a $K$-tier PPP network and we would like to be consistent in parameters.

Note that in (\ref{eq_MetaOurApp}), Lambert W function $W(\cdot,\cdot)$ is used. When we compute this function in MATLAB, we need to write the code by using logarithmic input which allows for much larger arguments than the built-in function to avoid infinity output, and related codes can be found in \cite{MATLAB:2010}.

Besides, to analyze the accuracy of the proposed approximation, we verify it through KL divergence \cite{csiszar1975divergence} numerically which compares the probability distribution of the proposed approximation with the exact distribution:
	\begin{align}
		D_{\rm KL,prop|exact} = \sum_{\gamma^{\prime} \in \chi} f_{1}^{\prime}(\gamma^{\prime},\theta) \log\bigg(\frac{f_{1}^{\prime}(\gamma^{\prime},\theta)}{f(\gamma^{\prime},\theta)}\bigg),\nonumber
	\end{align}
	in which $f_{1}^{\prime}(\gamma^{\prime},\theta)$ is obtained by discretizing $\bar{F}_{1}^{\prime}(\gamma,\theta)$: $f_{1}^{\prime}(\gamma^{\prime}(i),\theta) = \bar{F}_{1}^{\prime}(\gamma(i),\theta)-\bar{F}_{1}^{\prime}(\gamma(i+1),\theta)$, $\gamma^{\prime}(i) = \gamma(i)$, $\chi$ denotes the probability space: $\chi = 0.01,0.02,\cdots,0.99$, and similar steps for obtaining  $f(\gamma^{\prime},\theta)$ from $F(\gamma^{\prime},\theta)$. In addition, we compared the KL divergence of the proposed approximation with the beta approximation and the KL divergence of beta approximation is
	\begin{align}
		D_{\rm KL,beta|exact} = \sum_{\gamma^{\prime} \in \chi}f^{\prime}(\gamma^{\prime},\theta) \log\bigg(\frac{f^{\prime}(\gamma^{\prime},\theta)}{f(\gamma^{\prime},\theta)}\bigg),\nonumber
	\end{align}
	in which $f^{\prime}(\gamma^{\prime},\theta)$ is obtained by discretizing $\bar{F}^{\prime}(\gamma,\theta)$.
	Note that the above analysis is based on the numerical results, and the lower absolute value implies better approximation performance.

\begin{figure}
	\centering
	\includegraphics[width  = 1\columnwidth]{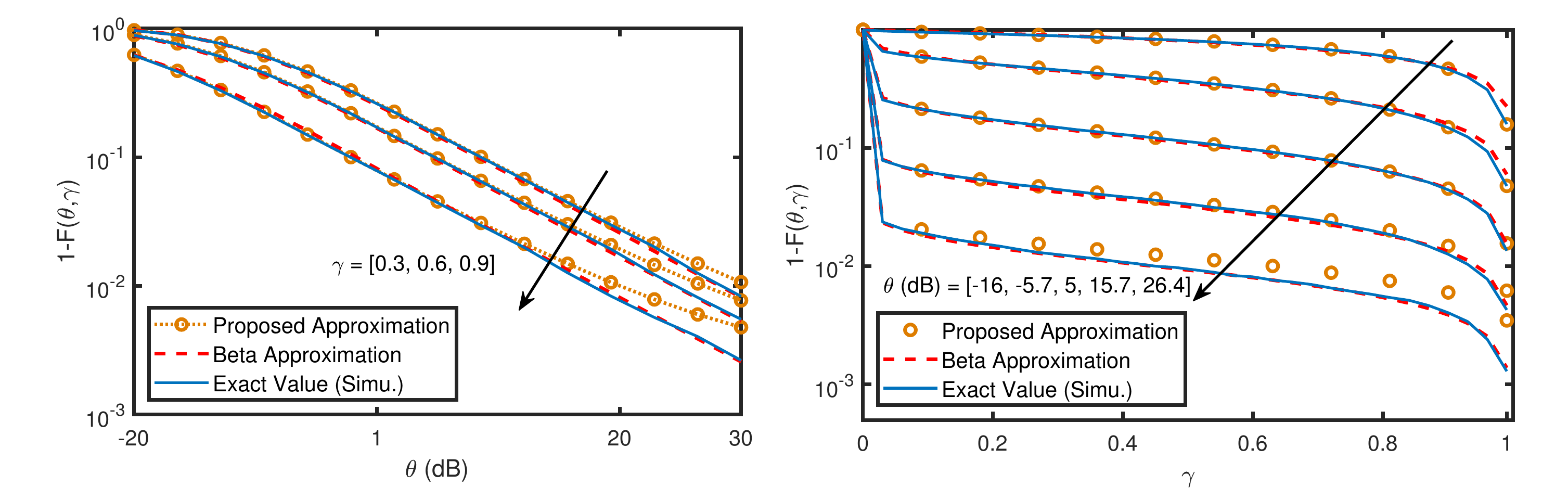}
	\caption{SINR meta distribution of PPP networks at $\lambda = 1$ BS/km$^2$, $\alpha = 4$ and $p_t = 10$ W. The solid lines are exact values based on simulations, dash lines are for beta approximation, and  markers for the proposed  approximation, respectively. }
	\label{fig_lambdab1}
\end{figure}
\begin{figure}
	\centering
	\includegraphics[width  = 1\columnwidth]{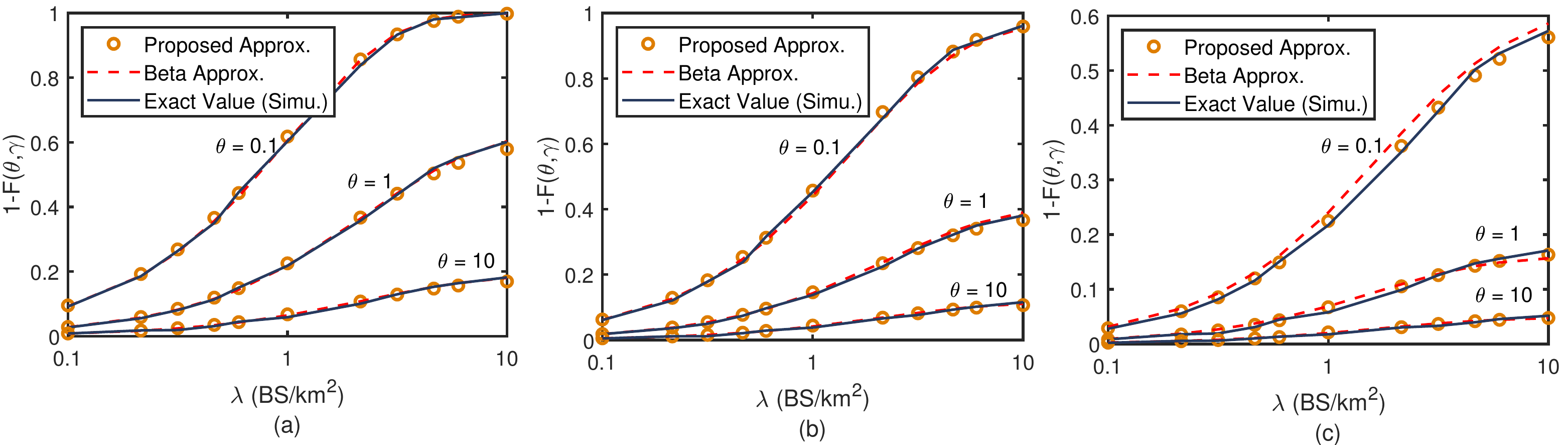}
	\caption{SINR meta distribution of PPP networks at different density values from $\lambda = 0.1$ to $\lambda = 10$ BS/km$^2$, $\alpha = 4$ and $p_t = 10$ W. The solid lines are exact values based on simulations, dash lines are for beta approximation and  markers are for the proposed  approximation, respectively. $\gamma = 0.3, 0.6, 0.9$ in \textbf{(a), (b), (c)}, respectively. }
	\label{fig_lambda_3}
\end{figure}
\begin{figure}
	\centering
	\subfigure[]{\includegraphics[width  = 0.49\columnwidth]{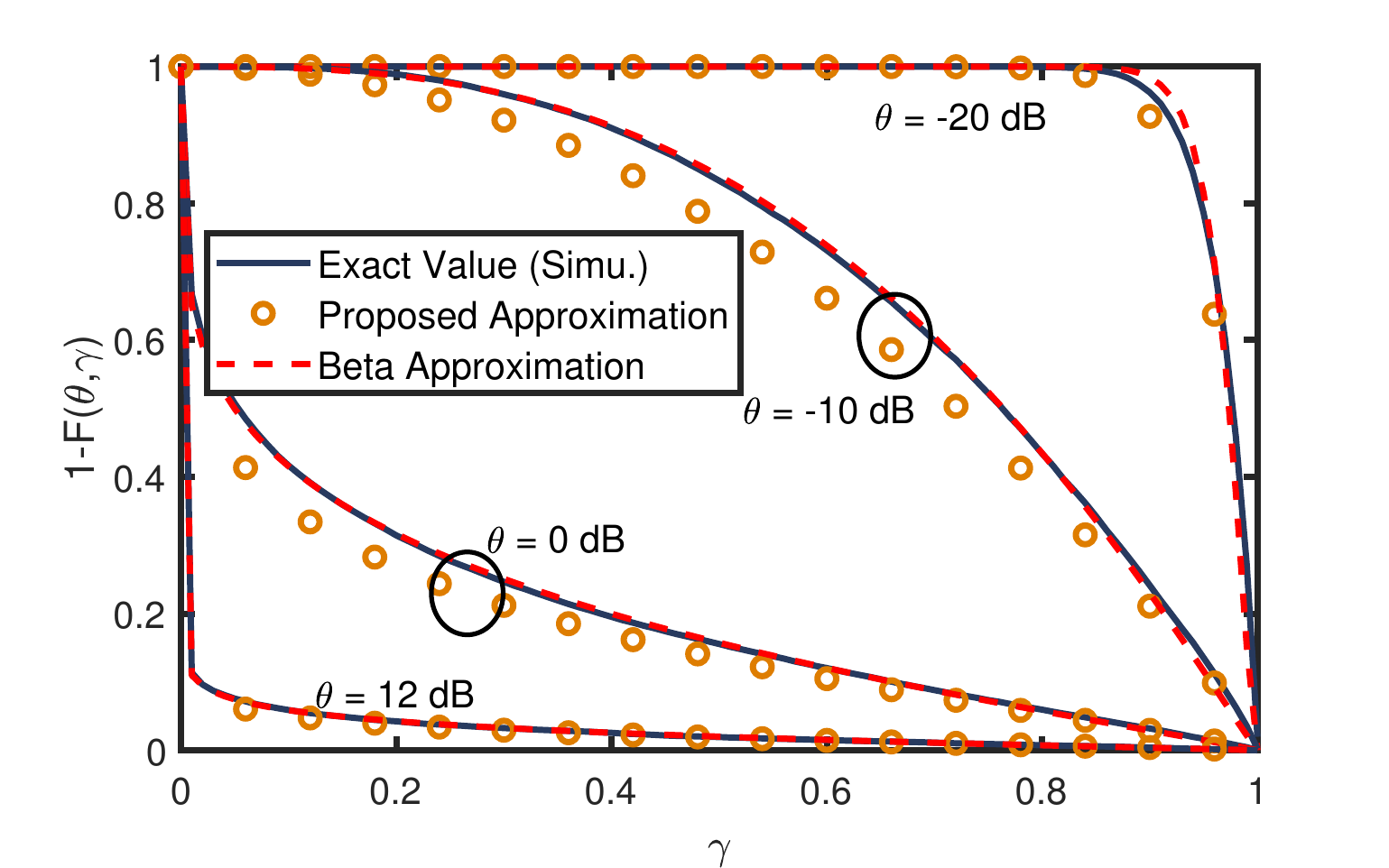}}
	\subfigure[]{\includegraphics[width  = 0.49\columnwidth]{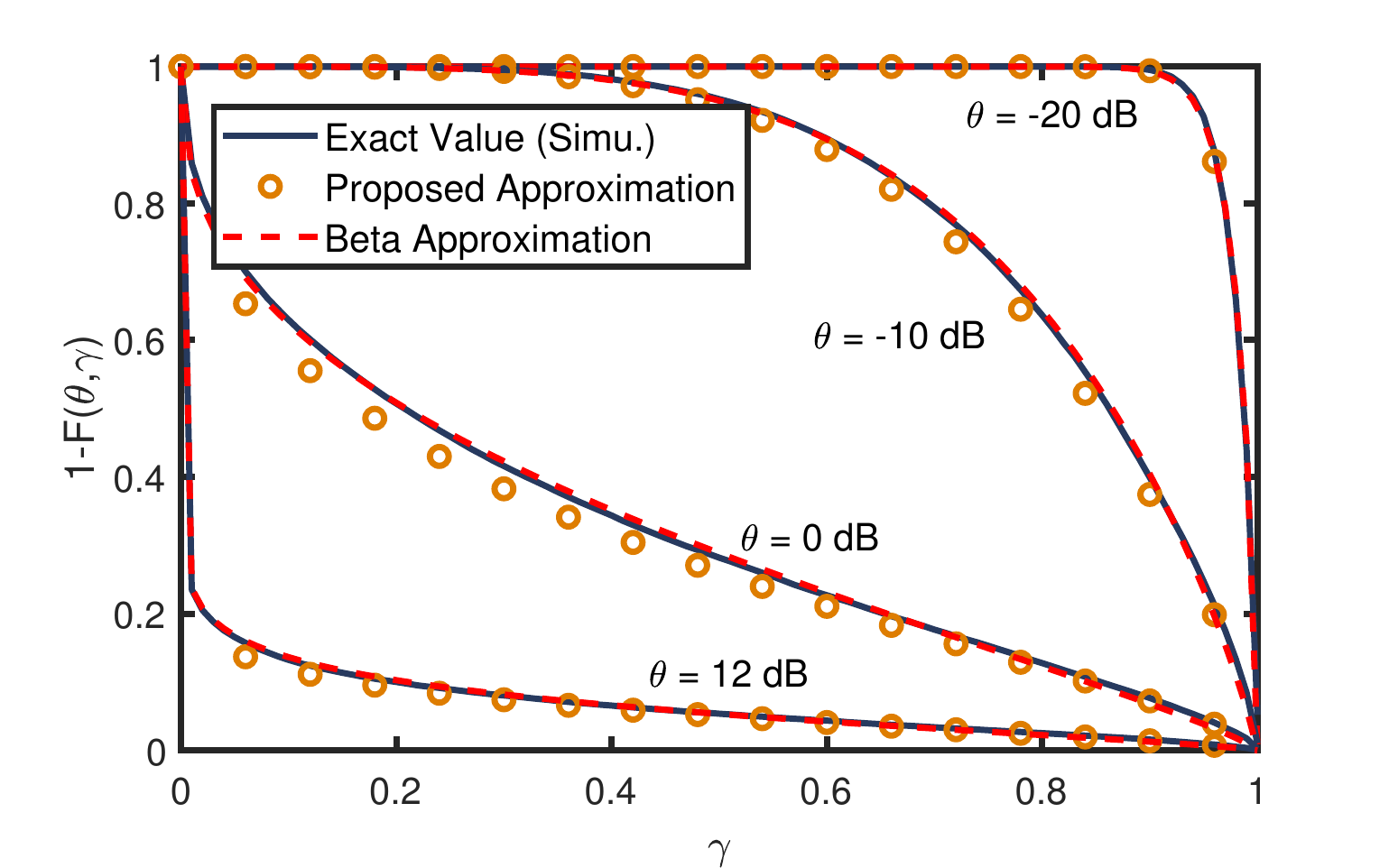}}
	\subfigure[]{\includegraphics[width  = 0.49\columnwidth]{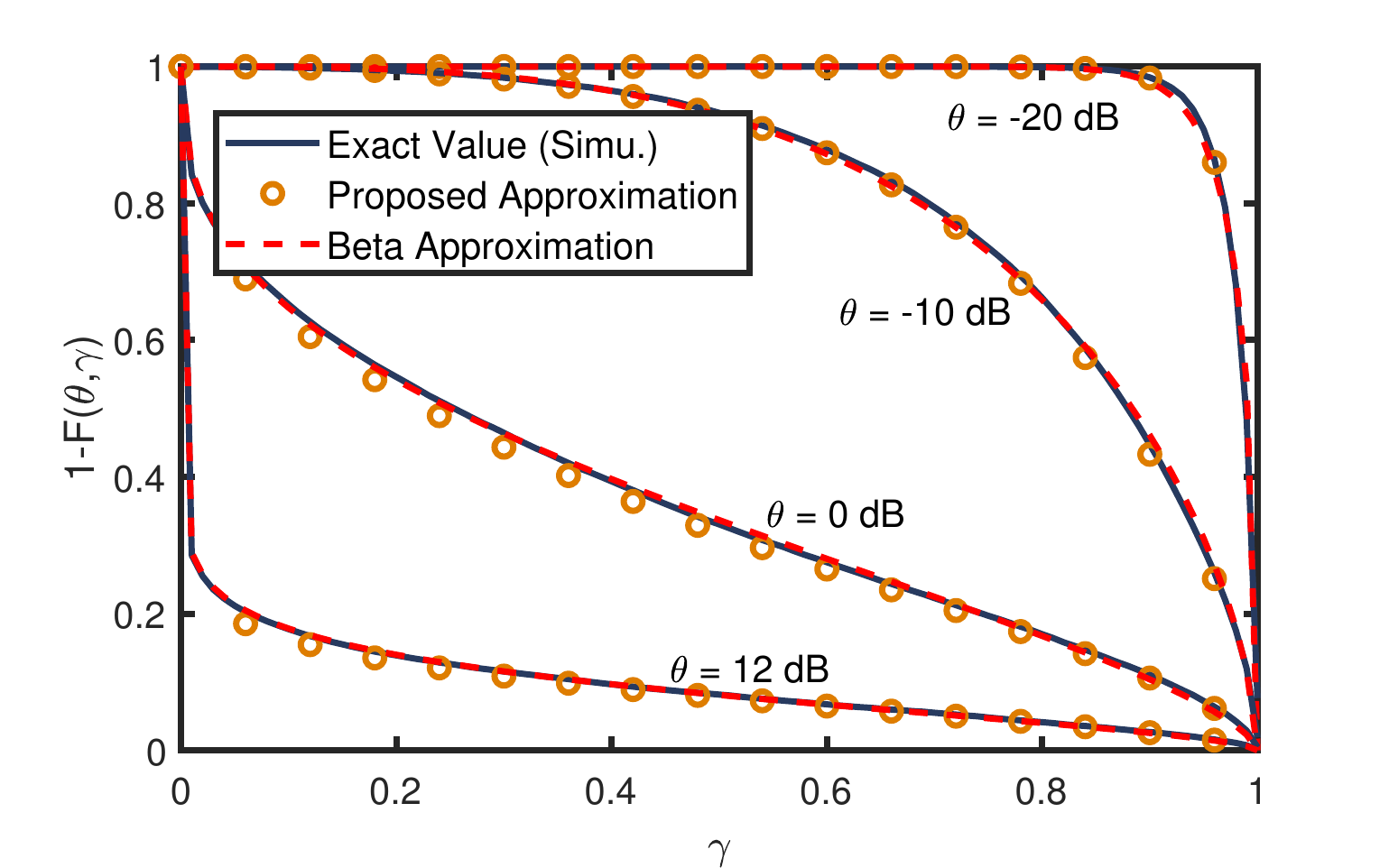}}	\subfigure[]{\includegraphics[width  = 0.49\columnwidth]{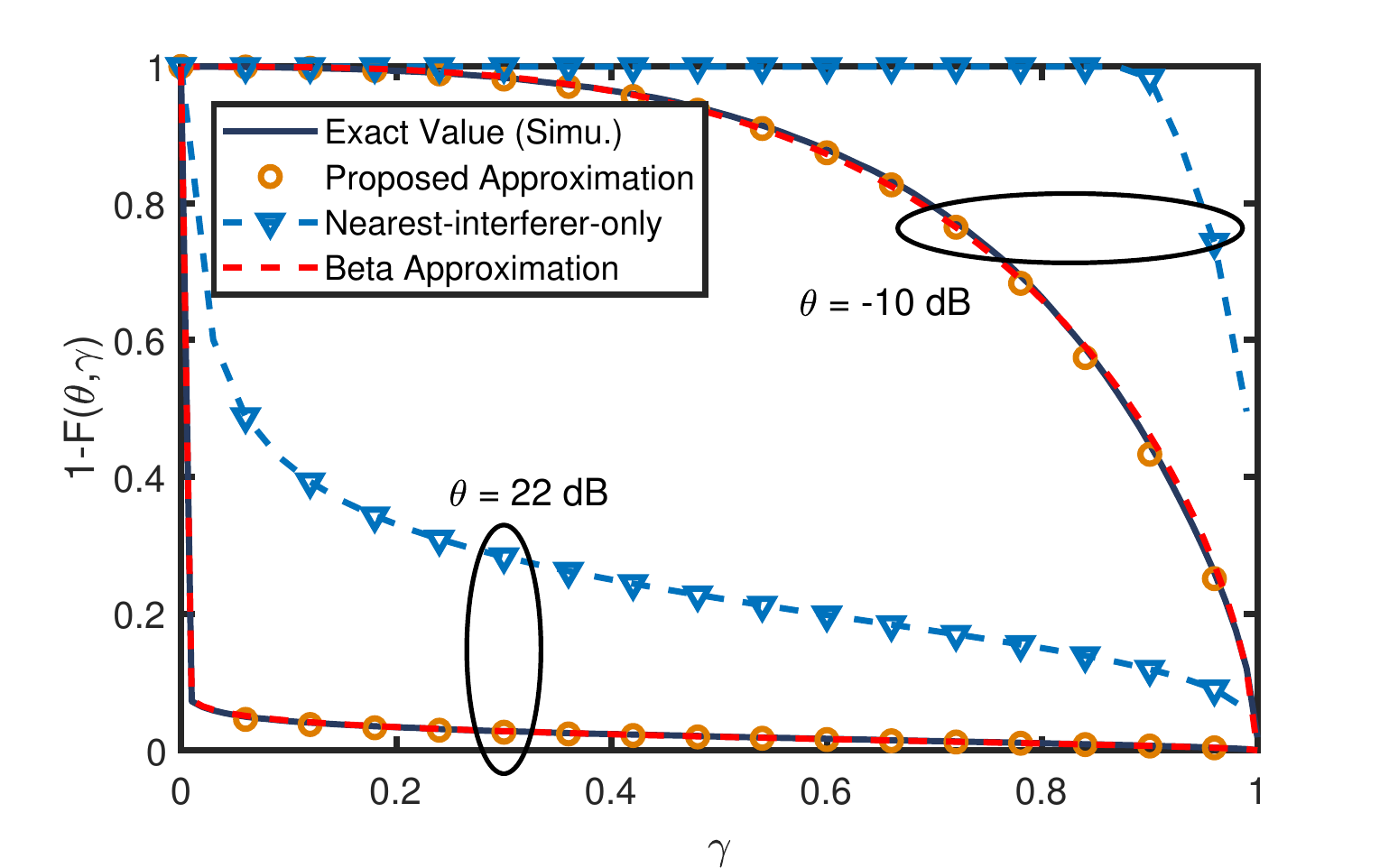}}
	\caption{SINR meta distribution of PPP networks at different path-loss values \textbf{(a)} $\alpha = 2.5$,  \textbf{(b)} $\alpha = 3$,  \textbf{(c)} $\alpha = 3.5$, respectively, while $\lambda = 1$ BS/km$^2$ and $p_t = 10$ W. The solid lines are exact values based on simulations, dash lines are for beta approximation and  markers are for the proposed  approximation, respectively. \textbf{(d)} The nearest-interferer-only approximation and the proposed dominant interferer-based approximation of SINR meta distribution at  $\alpha = 3.5$.}
	\label{fig_difalpha}
\end{figure}

In Fig.~\ref{fig_lambdab1}, we plot the exact value, beta approximation and  the proposed  approximation of SINR meta distribution against $\theta$ or $\gamma$ at different values of $\gamma$ or $\theta$ under $\lambda = 1$ km$^{-2}$. As shown,  the proposed  approximation shows good matching at low values of $\theta$ and $\gamma$. However, when it comes to very high values, especially $\theta$, a small gap exists.  The reason for the gap at high values of $\theta$ is that we ignore the high order terms of $\theta$ in both (\ref{eq_appPs}) and (\ref{eq_poof_MD}). Such gaps exist even if we increase the values of $j$ in (\ref{eq_j_appMeta}) ($j = 2$ shows almost the same curve as $j = 1$, in which the difference is less than $0.1$, hence, omitted in Fig. 5), hence the dominant interferer-based approximation (Cor. \ref{cor_dom}) is a more efficient way (comparing (\ref{eq_MetaOurApp}) to (\ref{eq_j_appMeta}) ) since no need of computing joint distribution of $R_1,\cdots,R_j$ (only $f_{R_1}(r)$ and $F_{R_0|R_1}(r)$ are required) and a simple integral completes the analysis. Notice that, the logarithmic scale is used for the y-axis in Fig.~\ref{fig_lambdab1} since the gap is negligible in  linear scale (e.g., the gap between  the proposed  approximation and exact value at $\theta = 26$ dB is actually 0.02).

Fig.~\ref{fig_lambda_3} shows the gap between  the proposed  approximation and beta and exact value of SINR meta distribution for a large range of BS densities. The proposed approximation shows very good performance at a large range of density of BS and low values of $\theta$, and its performance is even better than beta distribution, while saving a large amount of computing time.

Notice that the accuracy of dominant interferer-based approximation strongly depends on the path loss exponent. While $\alpha = 4$ is an important case, user connections in mm-Wave band will be line-of-sight, say UAV networks, hence in Fig.~\ref{fig_difalpha} (a),(b),(c) we plot $\alpha = 2.5,3,3.5$, respectively. Clearly, the proposed approximation becomes more accurate with the increase of $\alpha$ and in small values of $\alpha$ a gap exists. In Fig.~\ref{fig_difalpha} (d) we plot the nearest-interferer-only approximation, which is an upper bound of the system performance, and as mentioned in Remark \ref{rem_nearest}, the proposed approximation is tighter by considering the mean of the remaining interferers (all the interferers except the nearest one) but it is neither upper bound nor lower bound: as can be observed from Fig. \ref{fig_lambdab1} that the proposed approximation does indeed become higher than the exact value specially at high values of $\theta$. On the other hand, it can be observed from Fig. \ref{fig_difalpha} that the proposed approximation is lower than the exact value.

\begin{table}[ht]
	\centering
	\caption{KL divergence analysis of PPP networks}
	\label{KL_ppp}
	\begin{tabular}{l|l|l|l|l|l}
		\hline
	Parameters	& $\alpha = 2.5$ & $\alpha = 3$ & $\alpha = 3.5$ & $\alpha = 4$ & $\alpha = 4,\lambda = 0.1,0.5,5,10$ km$^{-2}$\\\hline
	$D_{\rm KL,prop|exact}$ ($\theta = -10$ dB) & 0.0417 &  0.0095& 0.0082  & 0.0037 & 0.0097,0.0050,0.0085,0.0119 \\
	$D_{\rm KL,beta|exact}$ ($\theta = -10$ dB)	& 0.0122  & 0.0284 & 0.0101 & -0.0045 & 0.0111,0.0025,0.0042,0.0335  \\\hline
	$D_{\rm KL,prop|exact}$ ($\theta = 0$ dB)	&0.0109 &  0.0105& 0.0037 & 0.0020 & 0.0007,0.0123,0.0039,0.0180 \\
	$D_{\rm KL,beta|exact}$ ($\theta = 0$ dB)	&0.0142 & 0.0126 & 0.0079 & 0.0064 &  0.0013,0.0167,0.0138,0.0322\\\hline
	$D_{\rm KL,prop|exact}$ ($\theta = 12$ dB)	& 0.0012& 0.006 & 0.0042 & 0.0011 & -0.0030,0.0033,0.0064,0.0146 \\
	$D_{\rm KL,beta|exact}$ ($\theta = 12$ dB)	&0.0031 & 0.0025 & 0.0017 & 0.0016 & -0.0025,0.0039,0.0033,0.0094 \\\hline
	\end{tabular}
\end{table}

The KL divergence analysis of the proposed approximation is provided in Table \ref{KL_ppp}. As shown in Table \ref{KL_ppp}, the proposed approximation at high values of $\alpha$ shows a competitive results compared with the beta approximation.

In the following part of this section, we follow the same simulation steps by conditioning on each realization of point processes, generating a large iterations of channel fading and computing the conditional success probability of each link. We plot the simulation results of the proposed approximation in the four different scenarios, Poisson bipolar network, MCP, $K$-tier PPP and PLCP, and compare the results with the traditional beta approximation and simulation based exact value of SINR meta distribution.

\subsection{Numerical Results of Poisson Bipolar Networks}
\begin{figure}[ht]
	\centering
	\includegraphics[width  = 1\columnwidth]{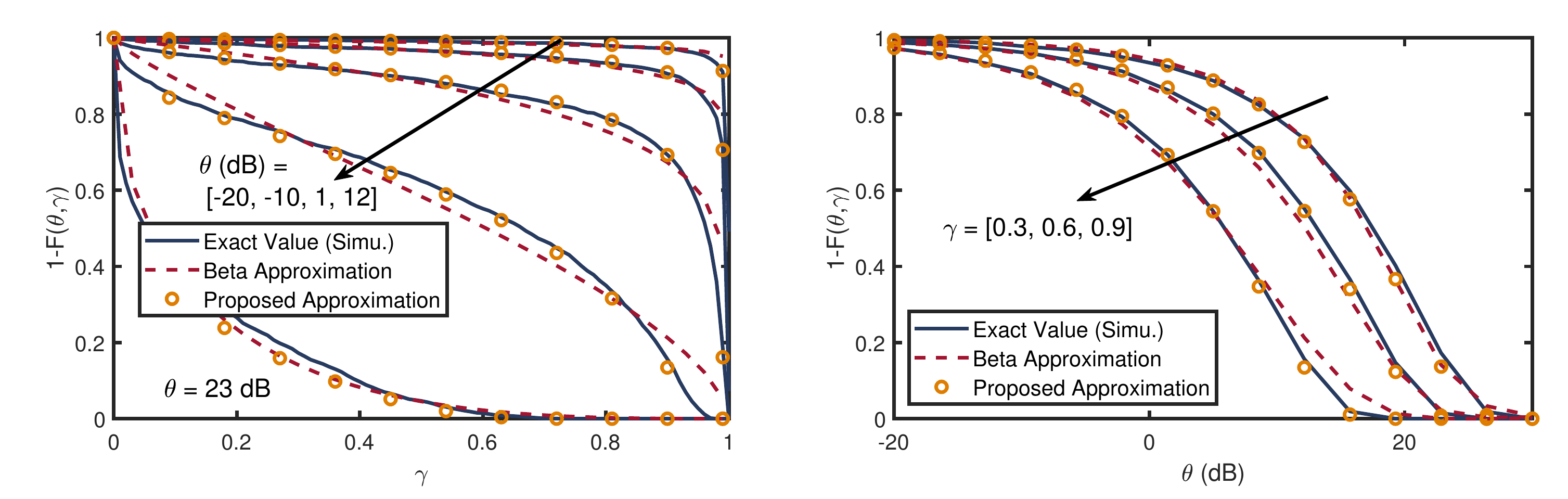}
	\caption{Meta distribution of a Poisson bipolar network at $\lambda = 10$ BS/km$^2$, $\alpha = 4$, $p_t = 10$ W and $R = 50$ m. The solid lines are exact values based on simulations and  markers for the proposed  approximation, respectively. }
		\label{fig_PB_1}
\end{figure}
\begin{figure}[ht]
	\centering
	\includegraphics[width  = 1\columnwidth]{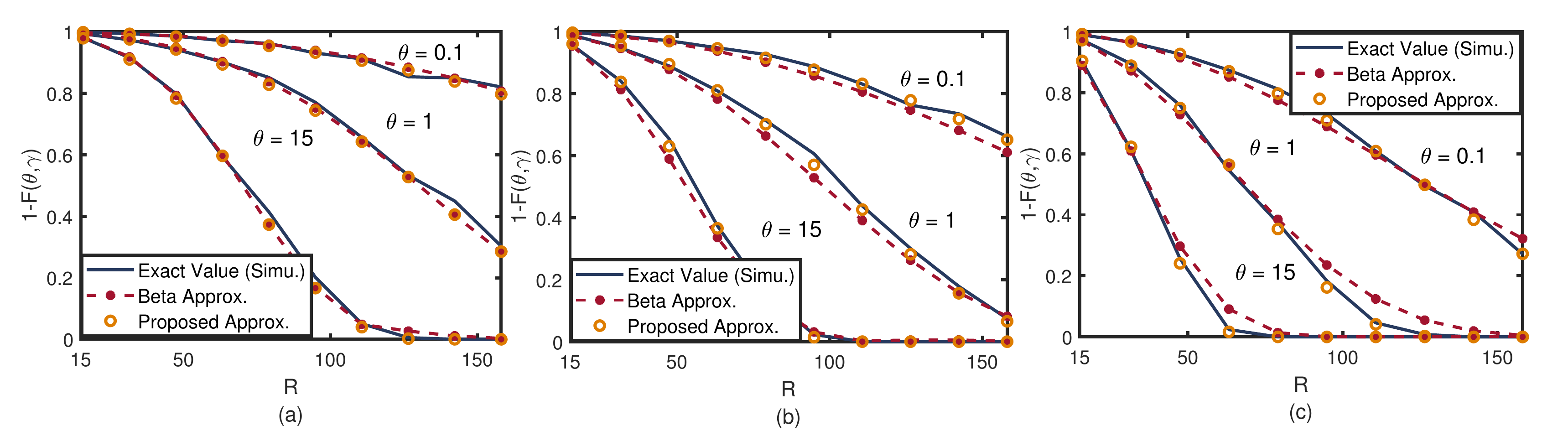}
	\caption{Meta distribution of a Poisson bipolar network at $\lambda = 10$ BS/km$^2$, $\alpha = 4$, $p_t = 10$ W. The solid lines are exact values based on simulations, dash lines are for beta approximation and  markers are for the proposed  approximation, respectively. $R$ is in meter and $\gamma = 0.3, 0.6, 0.9$ in \textbf{(a,b,c)}, respectively.  }
			\label{fig_PB_3}
\end{figure}
Fig. \ref{fig_PB_1} and Fig. \ref{fig_PB_3} show the SINR meta distribution of Poisson bipolar networks. The proposed approximation shows good matching at all values of $\theta$ and $\gamma$. In Fig. \ref{fig_PB_3}, we plot the SINR meta distribution under $10$ different values of $R$. With the increase of the distance, system's reliability drops sharply and approaches 0, which is owing to the fixed transmission distance and the noise power.

\begin{table}[ht]
	\centering
	\caption{KL divergence analysis of Poisson bipolar networks}
	\label{KL_PB}
	\begin{tabular}{l|l|l|l|l|l|l|l|l}
		\hline
		Parameters	& $R = 15$ & $R = 30$ & $R = 45$ & $R = 60$ & $R = 75$ & $R = 100$& $R = 125$& $R = 150$\\\hline
		$D_{\rm KL,prop|exact}$ ($\theta = -10$ dB)& 0.0013	& 0.0068 & 0.0241 & -0.0030 & 0.0367&0.0255 &0.0289&0.0127 \\
		$D_{\rm KL,beta|exact}$ ($\theta = -10$ dB)	&-0.0065& -0.0226 & -0.0285 & -0.0646 & -0.0814   &-0.1464&-0.1239&-0.0165\\\hline
		$D_{\rm KL,prop|exact}$ ($\theta = 0$ dB)&	-0.0034&  0.0006 & 0.0386 & 0.0269 &0.0153 &0.0098 &0.0096&0.0230\\
		$D_{\rm KL,beta|exact}$ ($\theta = 0$ dB)&	-0.0267& -0.0552 & -0.1085 & -0.1574 &-0.0740  &0.0173&0.0900&0.0164\\\hline
		$D_{\rm KL,prop|exact}$ ($\theta = 12$ dB)&	0.0142& 0.0440 & 0.0097 & 0.0148 &0.0186 &0.0207 &0.0204&0.0050\\
		$D_{\rm KL,beta|exact}$ ($\theta = 12$ dB)&	-0.0410& -0.1482 & -0.0015 & 0.0543 &0.0348  &0.0090&0.0379&0.0143\\\hline
	\end{tabular}
\end{table}

The KL divergence analysis of the proposed approximation  under Poisson bipolar networks is provided in Table \ref{KL_PB}, in which the unit of $R$ is meter. As shown in Table \ref{KL_PB}, the proposed approximation shows a competitive results compared with the beta approximation.

\subsection{Numerical Results of MCP}
\begin{figure}[ht]
	\centering
	\includegraphics[width  = 1\columnwidth]{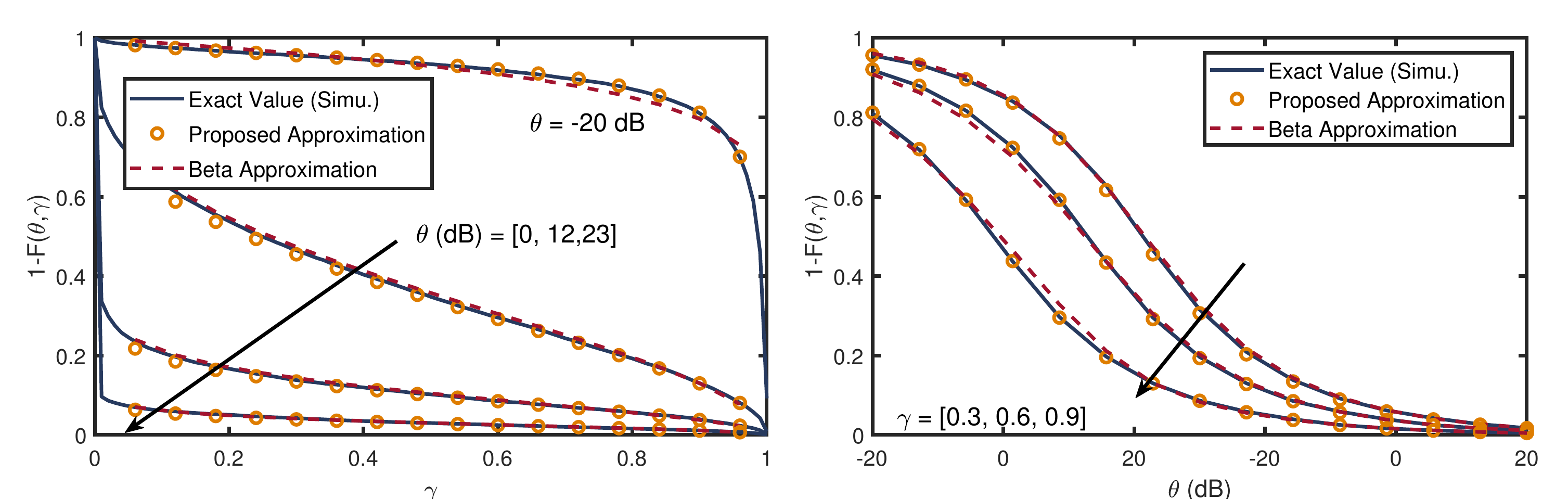}
	\caption{Meta distribution of MCP networks at $\lambda = 10$ BS/km$^2$, $\alpha = 4$ and $p_t = 10$ W. The solid lines are exact values based on simulations, dash lines are for beta approximation and  markers for the proposed  approximation, respectively. }
			\label{fig_MCP_1}
\end{figure}
\begin{figure}[ht]
	\centering
	\includegraphics[width  = 1\columnwidth]{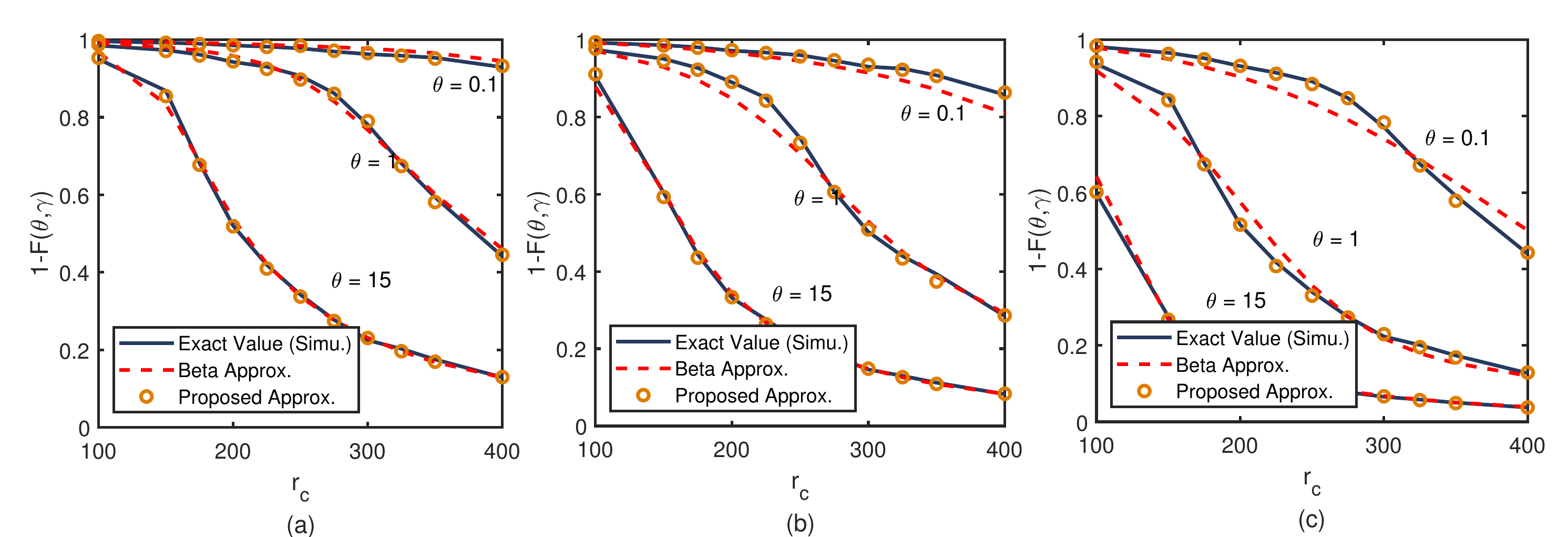}
	\caption{Meta distribution of a MCP network at $\lambda = 1$ BS/km$^2$, $\alpha = 4$ and $p_t = 10$ W. The solid lines are exact values based on simulations, dash lines are for beta approximation and  markers are for the proposed  approximation, respectively. $r_c$ is in meter and $\gamma = 0.3, 0.6, 0.9$ in \textbf{(a,b,c)}, respectively. }
				\label{fig_MCP_2}
\end{figure}
Fig. \ref{fig_MCP_1} and Fig. \ref{fig_MCP_2} show the SINR meta distribution of MCP networks. In Fig. \ref{fig_MCP_2}, we plot the SINR meta distribution under different values of user cluster radii. As expected, SINR meta distribution decrease with the increase of the user cluster radii. Besides,  compared to the Poisson bipolar networks, the reliability of MCP networks drops slower and does not approach 0. Clearly, the proposed approximation shows good performance, especially at large values of $\gamma$.

\begin{table}[ht]
	\centering
	\caption{KL divergence analysis of MCP networks}
	\label{KL_MCP}
	\begin{tabular}{l|l|l|l|l|l|l|l}
		\hline
		Parameters	& $r_c = 100$ & $r_c = 150$ & $r_c = 200$ & $r_c = 250$ & $r_c = 300$ & $r_c = 350$ & $r_c = 400$\\\hline
		$D_{\rm KL,prop|exact}$ ($\theta = -10$ dB) 	& -0.0031 & 0.0059 & -0.0015 &0.0173 & -0.0041&0.0218&-0.0012\\
		$D_{\rm KL,beta|exact}$ ($\theta = -10$ dB)	&  0.0008 & 0.0220 &0.0293 & 0.0249&-0.0345& -0.0161& -0.0116\\\hline
		$D_{\rm KL,prop|exact}$ ($\theta = 0$ dB)	& -0.0067 & 0.0117 & 0.0001 & 0.0063&-0.0027&0.0043&0.0006 \\
		$D_{\rm KL,beta|exact}$ ($\theta = 0$ dB)	& 0.0206 &0.0069  & -0.0237 & 0.0332&0.0343&0.0332&0.0136 \\\hline
		$D_{\rm KL,prop|exact}$ ($\theta = 12$ dB)	& 0.0007 &0.0052  & 0.0004 & 0.0011& 0.0002&0.0004&0.0002\\
		$D_{\rm KL,beta|exact}$ ($\theta = 12$ dB)	& -0.0378 & 0.0435 & 0.0296 &0.0029 &-0.0011&-0.0018&-0.0115 \\\hline
	\end{tabular}
\end{table}

The KL divergence analysis of the proposed approximation  under MCP networks is provided in Table \ref{KL_MCP}, in which the unit of $r_c$ is meter. As shown in Table \ref{KL_MCP}, the proposed approximation shows a competitive results compared with the beta approximation.

\subsection{Numerical Results of $K$-tier Networks}
\begin{figure}[ht]
	\centering
	\includegraphics[width  = 1\columnwidth]{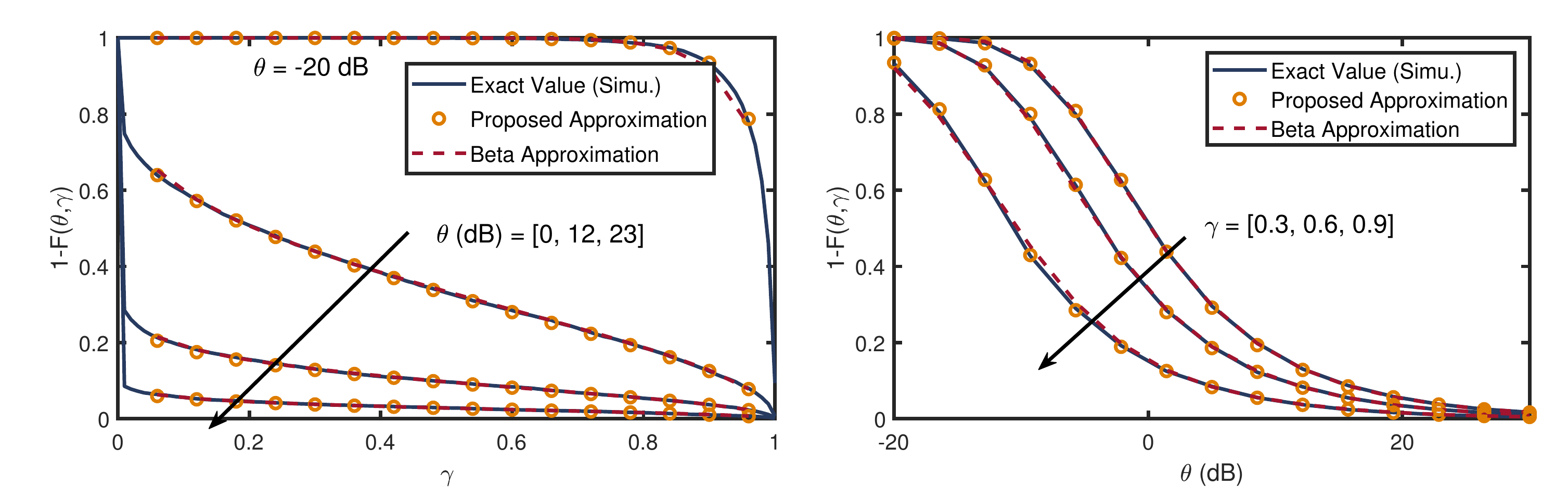}
	\caption{Meta distribution of a $K$-tier PPP network, where $K = 2$, $\lambda_{1} = 1$ BS/km$^2$, $\lambda_{2} = 3$ BS/km$^2$, $\alpha = 4$, $p_{t,1} = 10$ W and $p_{t,2} = 5$. The solid lines are exact values based on simulations, dash lines are for beta approximation and  markers for the proposed  approximation, respectively. }
				\label{fig_KPPP_1}
\end{figure}
\begin{figure}[ht]
	\centering
	\includegraphics[width  = 1\columnwidth]{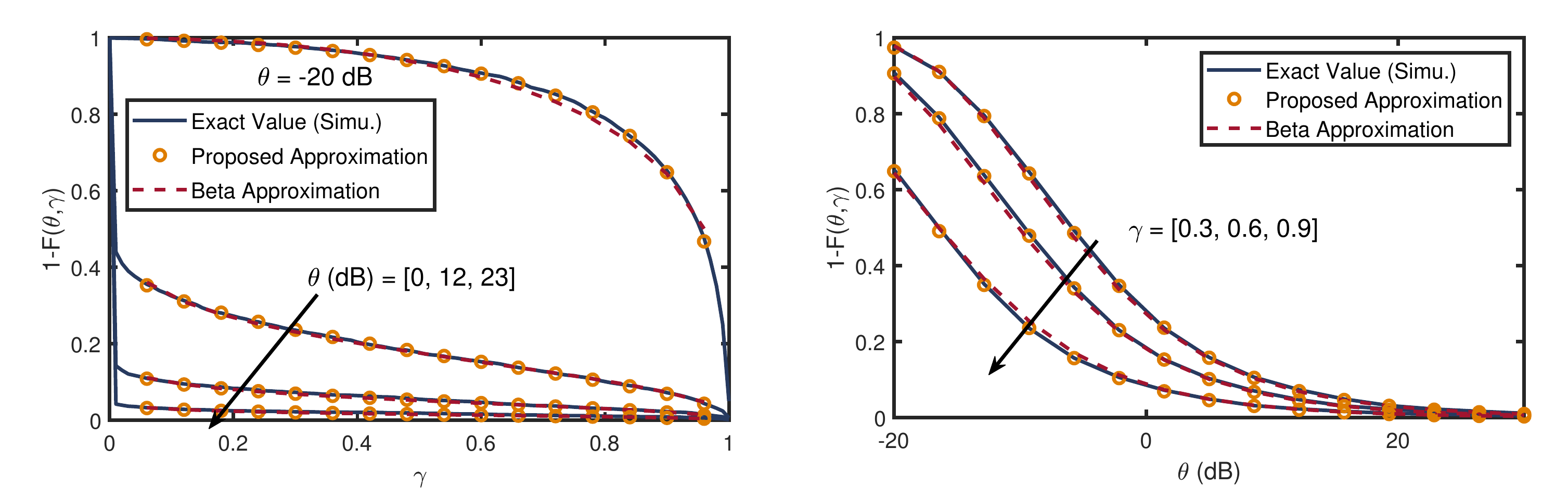}
	\caption{Meta distribution of a $K$-tier PPP network, where $K = 3$, $\lambda_{1} = 0.1$ BS/km$^2$, $\lambda_{2} = 0.5$ BS/km$^2$, $\lambda_{3} = 5$ BS/km$^2$, $\alpha = 4$, $p_{t,1} = 10$ W, $p_{t,2} = 2$ W and $p_{t,3} = 0.2$ W. The solid lines are exact values based on simulations, dash lines are for beta approximation and  markers for the proposed  approximation, respectively. }
	\label{fig_KPPP_2}
\end{figure}
Fig. \ref{fig_KPPP_1} and Fig. \ref{fig_KPPP_2} show the SINR meta distributions for a  $2$-tier and  $3$-tier network, respectively. Here we can see that the proposed approximation also works well for multi-tier networks by mapping the $i$-th tier to the first tier and obtaining the equivalent distance distributions and the density of the new point process. In this way, we can avoid computing the association probability and the summation of the aggregate interference of different tiers.

Since the analysis of $K$-tier PPP is similar to PPP, the KL divergence analysis is omitted.

\subsection{Numerical Results of PLCP}
\begin{figure}[ht]
	\centering
	\includegraphics[width  = 1\columnwidth]{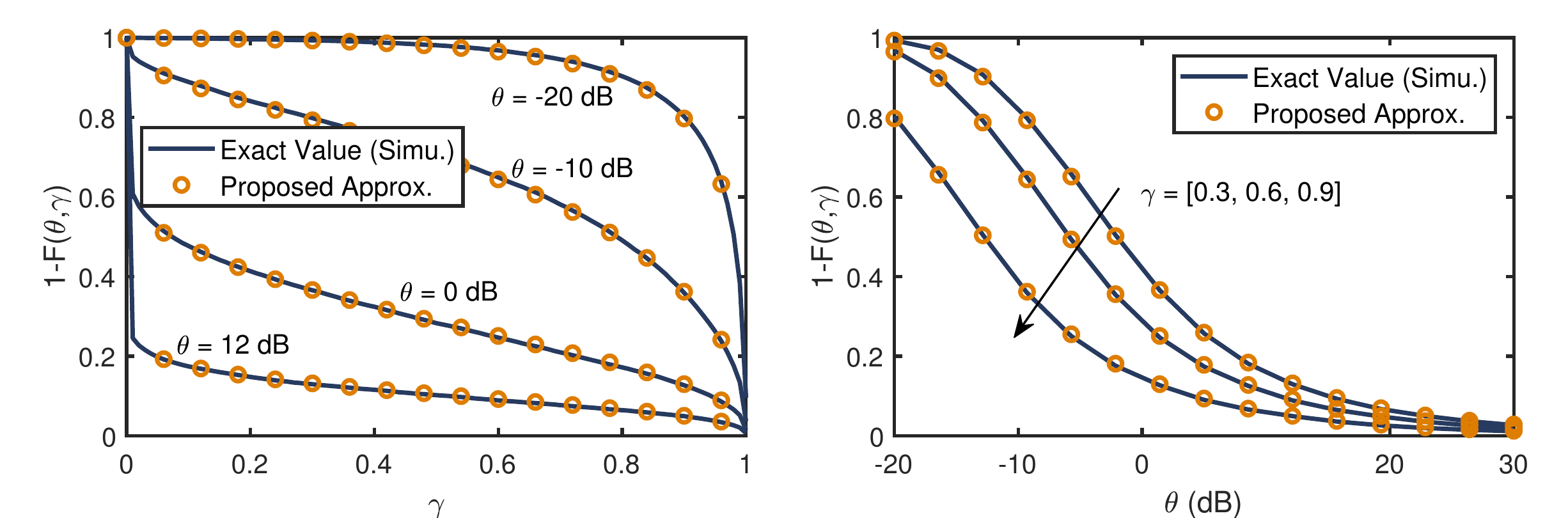}
	\caption{The SINR meta distribution of a PLCP network, where $\lambda_{l} = 8/\pi$ km/km$^2$, $\lambda_{p} = 0.2$ /km (the density of BSs is $1.6$ BS/km$^2$), $\alpha = 4$, $p_{t} = 10$ W. The solid lines are exact values based on simulations and  markers for the proposed  approximation, respectively. }
	\label{fig_PLCP_1}
\end{figure}
\begin{figure}[ht]
	\centering
	\includegraphics[width  = 1\columnwidth]{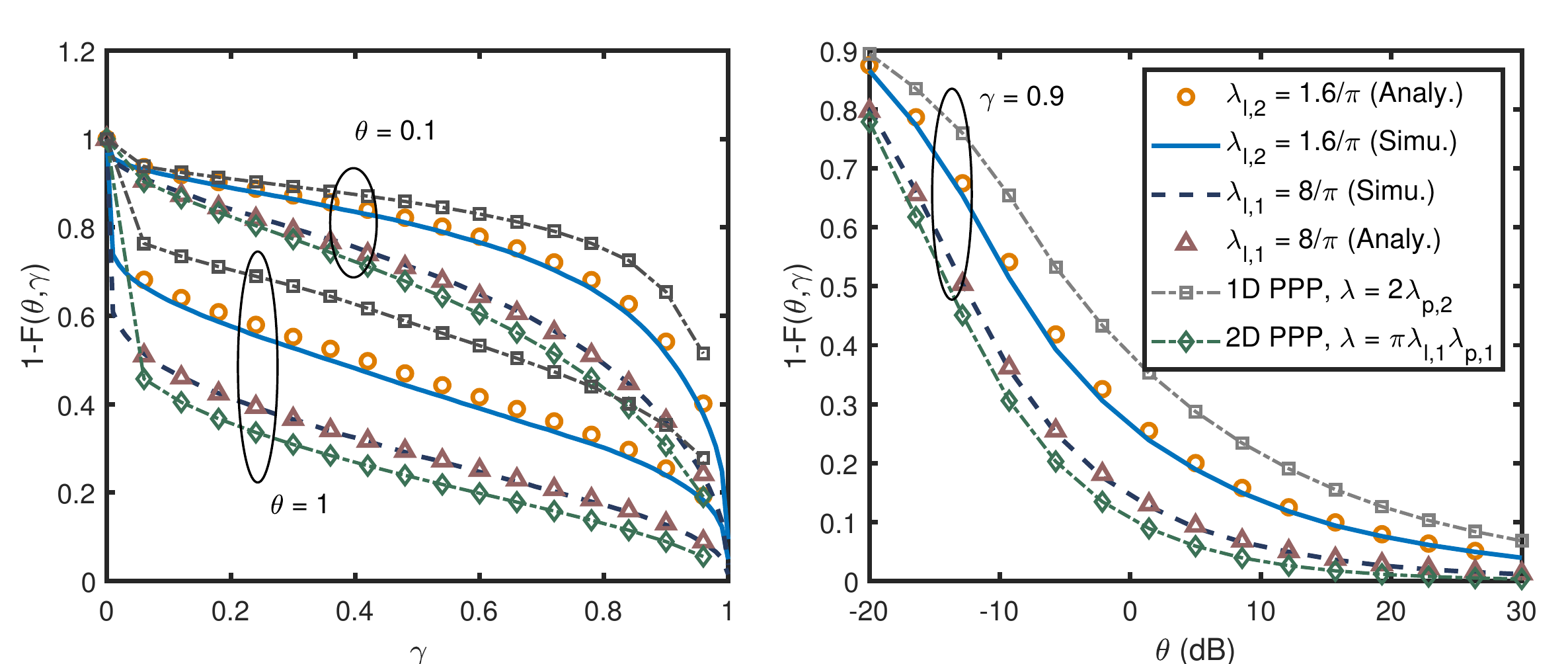}
	\caption{The SINR meta distribution of PLCP networks, while the density of the BS is fixed at $1.6$ BS/km$^2$, the line and point densities are different, $\lambda_{l,1} = 5\lambda_{l,2}$, $\alpha = 4$, $p_{t} = 10$ W. The unit of $\lambda_l$ is in km/km$^2$. The solid lines are exact values based on simulations and  markers for the proposed  approximation, respectively. The green curves with diamond markers are the SINR meta distribution of a 2D PPP network with density $\lambda = \pi\lambda_{l,1}\lambda_{p,1}$ and the gray curves with square markers are the SINR meta distribution of a 1D PPP network with density $\lambda = 2\lambda_{p,2}$.}
\label{fig_PLCP_2}
\end{figure}
Fig. \ref{fig_PLCP_1} and Fig.  \ref{fig_PLCP_2} display the SINR meta distribution of PLCP networks. In Fig. \ref{fig_PLCP_2}, we plot the SINR meta distribution in two different line and point densities. While we fix the overall density of the BSs, with the increasing of the line density, the fraction of links exceeding the reliability threshold increases. The reason is that as the line density increases ($\lambda_l \to \infty$) while the overall average number of points remains unchanged, the PLCP converges to a homogeneous 2D PPP with point density $\lambda = \pi\lambda_l\lambda_p$, as shown in the green curves with diamond markers in Fig. \ref{fig_PLCP_2}; and as the line density decreases ($\lambda_l \to 0$) while the overall average number of points remains unchanged, the PLCP reduces to a homogeneous 1D PPP with point density $\lambda = \lambda_p$. However, since the analysis of 1D PPP with $\lambda_p = \infty$ is challenging, we plot the SINR meta distribution of 1D PPP with density $\lambda = 2\lambda_{p,2}$, as shown in the gray curves  with square markers in Fig. \ref{fig_PLCP_2}. In these two asymptotical scenarios, the first contact distance in the case of 2D PPP is further than that of 1D PPP. Hence, system is more reliable with the line density decreasing since shorter serving distance and the system performance is bounded by 2D PPP and 1D PPP with corresponding point densities.

\begin{figure}[ht]
	\centering
	\includegraphics[width  = 0.8\columnwidth]{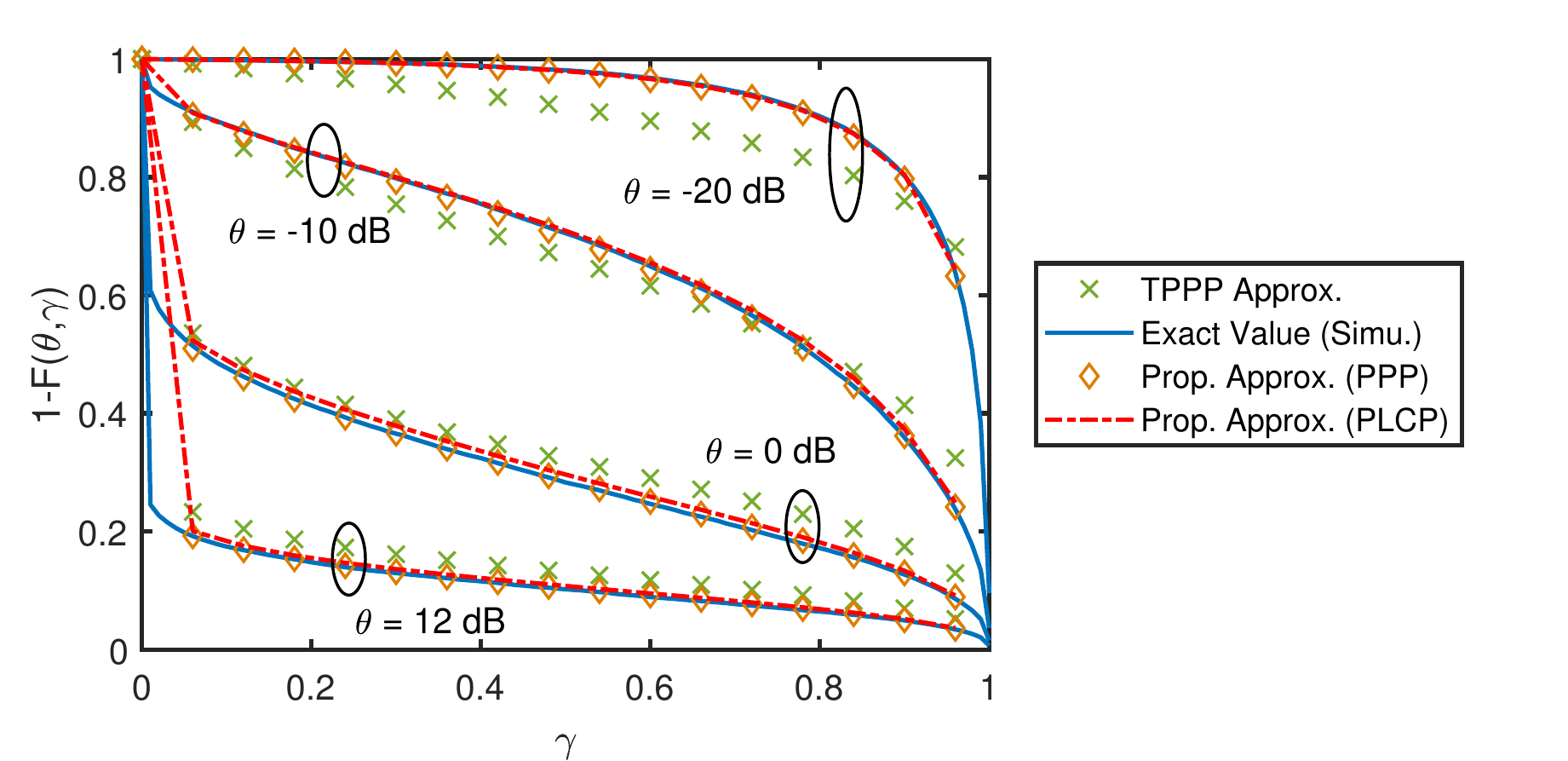}
	\caption{The SINR meta distribution of PLCP networks under three approximations, (i) TPPP approximation \cite[(16)]{9462466}: green and crossing markers, (ii) proposed approximation with PPP-approximated interference: orange and diamond markers, and (iii) proposed approximation with PLCP interference: red dash lines, and (iv) the exact value: blue and solid lines. The curves are plotted under $\lambda_l = 8/\pi$ km/km$^2$ and $\lambda_p = 0.2$ /km.}
	\label{fig_PLCP_comp}
\end{figure}

Besides, in Fig. \ref{fig_PLCP_comp}, we plot the proposed approximation with PLCP interference, PPP-approximated interference, respectively, and the approximation mentioned in \cite[(16)]{9462466}, which is based on the transdimensional PPP (TPPP). As shown, PPP is a good approximation of PLCP and all approximations provide good matching results.

\begin{table}[ht]
	\centering
	\caption{KL divergence analysis of PLCP networks}
	\label{KL_PLCP}
	\begin{tabular}{l|l|l}
		\hline
		Parameters	& $\lambda_l = 1.6/\pi$ km/km$^2$, $\lambda_p =1$ /km&  $\lambda_l = 8/\pi$ km/km$^2$,$\lambda_p = 0.2$ /km\\\hline
		$D_{\rm KL,prop|exact}$ ($\theta = -10$ dB) 	& -0.0234 &  -0.0031\\
		$D_{\rm KL,prop(PPP)|exact}$ ($\theta = -10$ dB)	& -0.0280 &  -0.0103  \\		$D_{\rm KL,TPPP(16)|exact}$ ($\theta = -10$ dB)	& 0.1796 &  -0.0606  \\\hline
		$D_{\rm KL,prop|exact}$ ($\theta = 0$ dB)	& -0.0079 & -0.0028 \\
		$D_{\rm KL,prop(PPP)|exact}$ ($\theta = 0$ dB)	& -0.0094 &  -0.0052 \\
		$D_{\rm KL,TPPP(16)|exact}$ ($\theta = 0$ dB)	& 0.0262 &  -0.0404 \\\hline
		$D_{\rm KL,prop|exact}$ ($\theta = 12$ dB)	& -0.0045 &  -0.0013 \\
		$D_{\rm KL,prop(PPP)|exact}$ ($\theta = 12$ dB)	& -0.0052 & -0.0020 \\		
		$D_{\rm KL,TPPP(16)|exact}$ ($\theta = 12$ dB)	& -0.0115 & -0.0143 \\\hline
	\end{tabular}
\end{table}

The KL divergence analysis of the proposed approximation  under PLCP networks is provided in Table \ref{KL_PLCP}. As shown in Table \ref{KL_PLCP}, the proposed approximation shows a good matching result, and the PPP interference also shows a good matching result, which is because PPP is a good approximation of PLCP.

\section{Conclusion}
This paper analyzes a dominant interferer-based approximation, considering the dominate interferers exactly while the rest interferers in an average sense, for SINR meta distribution in the downlink Poisson cellular networks. We first obtain the proposed approximation in a standard PPP networks with all the channel links subject to the Rayleigh fading. The applied approximation shows good matching with the exact value of SINR meta distribution at large range of BS densities, SINR thresholds and various system parameters. Compared with the traditional method, the proposed  approximation does not require computing the moments of conditional success probability. It can be derived in a simple form based on the Lambert W function and the joint CDF and PDF of the first and the second nearest BSs' distance distribution, which highly reduces calculation complexity and is highly time-efficient. To illustrate the accuracy and operability, we extend the proposed approximation in four different scenarios and compare the results with the  popular used beta approximation as well as the Monte-Carlo simulations. Meanwhile, we derive the SINR meta distribution for the PLCP networks for the first time based on the proposed approximation since it avoids computing the $b$-th moments and highly reduce the complexity of the calculation process.
\medskip

Throughout this paper, we focus on validating the proposed approximation with simulations and beta approximation in downlink scenarios. For the uplink cases, system models are more complex, e.g., transmit power may be a function of the serving distance (inverse power control), the locations of interferers can be closer to the BSs, and maybe more than one active user within one Voronoi cell (depends on the transmission policy). Therefore, the distance distributions, as well as system model, are way more complicated than downlink scenarios. While we only compute the SINR meta distribution in the downlink in this paper, uplink analysis, as well as some more complex system models, such as cluster distributed BSs \cite{saha2020meta} and \cite{shi2021meta}, can be an interesting future research direction.

\appendix

\subsection{Proof of Theorem \ref{theorem_exactmeta}} \label{app_exactmeta}
Given $\Phi$, the success probability is given by
	\begin{align}
		P_s(\theta) = \prod_{i\in\mathbb{N}}\bigg(\frac{1}{1+\theta ||x_i||^{-\alpha}||x_0||^{\alpha}}\bigg)\exp\bigg(-\frac{\theta}{p_t}||x_0||^{\alpha}\sigma^2\bigg).\label{eq_ps}
	\end{align}
	Hence, following the Campbell's  theorem \cite{haenggi2012stochastic} with conversion from Cartesian to polar coordinates, we have the $b$-th  moment,
	\begin{align}
		&M_{b}(\theta) =  \mathbb{E}_{||x_0||}\bigg[\mathbb{E}\bigg[\prod_{i\in\mathbb{N}}\bigg(\frac{1}{1+\theta ||x_i||^{-\alpha}||x_0||^{\alpha}}\bigg)^b\biggm\vert||x_0||\bigg]\exp\bigg(-\frac{b\theta}{p_t}||x_0||^{\alpha}\sigma^2\bigg)\bigg]\nonumber\\
		&= \mathbb{E}_{R_0}\bigg[\exp\bigg(-2\pi\lambda\int_{r_0}^{\infty}\bigg[1-\bigg(\frac{1}{1+\theta r^{-\alpha}R_{0}^{\alpha}}\bigg)^b\bigg]r{\rm d}r\bigg)\exp\bigg(-\frac{b\theta}{p_t}R_0^{\alpha}\sigma^2\bigg)\bigg]\nonumber\\
		&= \mathbb{E}_{R_0}\bigg[\exp\bigg(-\lambda F_b \bigg)\exp\bigg(-\frac{b\theta}{p_t}R_0^{\alpha}\sigma^2\bigg)\bigg],
	\end{align}
	where,
\begin{align}
	&F_b = 2\pi\int_{R_0}^{\infty}\bigg[1-\bigg(\frac{1}{1+\theta r^{-\alpha}R_{0}^{\alpha}}\bigg)^b\bigg]r{\rm d}r\stackrel{(a)}{=} \pi\delta\int_{R_{0}^{\alpha}}^{\infty}\bigg[1-\bigg(\frac{u}{u+\theta R_{0}^{\alpha}}\bigg)^b\bigg]u^{\delta-1}{\rm d}u\nonumber\\
	&\stackrel{(b)}{=} \pi\delta\int_{R_{0}^{\alpha}}^{\infty}\sum_{k=1}^{\infty}\binom{b}{k}(-1)^{k+1}\bigg(\frac{\theta R_{0}^{\alpha}}{u+\theta R_{0}^{\alpha}}\bigg)^k u^{\delta-1}{\rm d}u= \pi\delta\sum_{k=1}^{\infty}\binom{b}{k}(-1)^{k+1}(\theta R_{0}^{\alpha})^k \int_{R_{0}^{\alpha}}^{\infty}\bigg(\frac{u^{\delta-1}}{(u+\theta R_{0}^{\alpha})^k}\bigg) {\rm d}u\nonumber\\
	&= \pi\delta\sum_{k=1}^{\infty}\binom{b}{k}(-1)^{k+1} \frac{\alpha \theta^k R_0^{2}}{-2+k\alpha}{}_{2}F_{1}(k,-\delta+k,1-\delta+k,-\theta),
\end{align}
where step $(a)$ follows from using the replacement $u = r^\alpha$ and step $(b)$ follows from the binomial expansion. Notice that the above expression is different from the one derived in \cite{haenggi2013diversity}, since the association policies are different: the typical user in this section associated with the nearest BS, while the user in the Poisson bipolar network has a dedicated serving BS. Therefore, the locations of interfering BSs in this section are different from \cite{haenggi2013diversity}: lower bound of the integration is not zero. Besides, since we included the noise here, the relative distance process (RDP) approach mentioned in \cite{haenggi2015meta} and \cite{ganti2015asymptotics} cannot be used.
\bibliographystyle{IEEEtran}
\bibliography{Ref7}
\end{document}

%% file: notation.tex
\def\nb0{{\mathbf{0}}}
\def\nb1{{\mathbf{1}}}

\newtheorem{lemma}{Lemma}

\newtheorem{definition}{Definition}

\newtheorem{theorem}{Theorem}

\newtheorem{cor}{Corollary}

\newtheorem{remark}{Remark}

